\documentclass[11pt,a4paper]{article}
\pdfoutput=1
\usepackage{jcappub}

\usepackage{amsmath}
\usepackage{amssymb}  
\usepackage[english]{babel} 
\usepackage{booktabs} 
\usepackage{color} 
\usepackage{dcolumn}
\usepackage{pdflscape}
\usepackage[T1]{fontenc}
\usepackage{graphicx}

\newcommand{\aJLA}{\alpha_\mathrm{JLA}}
\newcommand{\bJLA}{\beta_\mathrm{JLA}}

\newcommand{\abs}[1]{\lvert #1 \rvert}

\newcommand{\be}{\begin{equation}}
\newcommand{\ee}{\end{equation}}

\newcommand{\MN}{\textsc{MultiNest}}

\newcommand{\aanda}{Astron. Astrophys.}
\newcommand{\aipcp}{AIP Conf. Proc.}
\newcommand{\apj}{Astrophys. J.}
\newcommand{\apjs}{Astrophys. J. Suppl.}

\newcommand{\mnras}{Mon. Not. Roy. Astron. Soc.}
\newcommand{\prd}{Phys. Rev. D}

\newcolumntype{d}[1]{D{.}{.}{#1}}
\newcolumntype{v}[1]{D{,}{,\ }{#1}}

\begin{document}
\title{Bayesian Comparison of Interacting Scenarios}
\author[a,b]{Antonella Cid}
\author[a]{Beethoven Santos}
\author[c]{Cassio Pigozzo}
\author[d]{Tassia Ferreira}
\author[a,e]{Jailson Alcaniz}
		
\affiliation[a]{\small\it Observat\'orio Nacional, 20921-400, Rio de Janeiro, RJ, Brasil.}
\affiliation[b]{\small\it Departamento de F\'isica, Grupo Cosmolog\'ia y Part\'iculas Elementales, Universidad del B\'io-B\'io, Casilla 5-C, Concepci\'on, Chile.}
\affiliation[c]{\small\it Instituto de F\'isica, Universidade Federal da Bahia, 40210-340, Salvador, BA, Brasil.}
\affiliation[d]{\small\it PPGCosmo, CCE, Universidade Federal do Esp\'irito Santo, 29075-910, Vit\'oria, ES, Brasil.}
\affiliation[e]{\small\it Departamento de F\'isica, Universidade Federal do Rio Grande do Norte, 59072-970, Natal, RN, Brasil }

\emailAdd{acidm@ubiobio.cl}
\emailAdd{thoven@on.br}
\emailAdd{cpigozzo@ufba.br}	
\emailAdd{t.ferreira@cosmo-ufes.org}
\emailAdd{alcaniz@on.br}

\date{\today}
\abstract{
We perform a Bayesian model selection analysis for different classes of phenomenological coupled scenarios of dark matter and dark energy with linear and non-linear interacting terms. We use a combination of some of the latest cosmological data such as type Ia supernovae (SNe Ia),  cosmic chronometers (CC), cosmic microwave background (CMB) and two sets of baryon acoustic oscillations measurements, namely, 2-dimensional angular measurements (BAO2) and 3-dimensional angle-averaged measurements (BAO3).
We find weak and moderate evidence against two-thirds of the interacting scenarios considered with respect to $\Lambda$CDM when the full joint analysis is considered. About one-third of the models provide a description to the data as good as the one provided by the standard model. 
Our results also indicate that either SNe Ia, CC or BAO2 data by themselves are not able to distinguish among interacting models or $\Lambda$CDM but the standard BAO3 measurements and the combination with the CMB data are indeed able to discriminate among them. We find that evidence disfavoring interacting models is weaker when we use BAO2 (data claimed to be almost model-independent) instead of the standard BAO3 measurements. These results help select classes of viable and non-viable interacting models in light of current data.}
	
\maketitle

\section{Introduction}
In the coming decade cosmological data from a number of planned galaxy surveys and cosmic microwave background experiments will be used to tackle fundamental questions about the nature of the physical make-up of the Universe. Currently, the standard picture used to describe the available observations is the $\Lambda$CDM cosmology,  
a model in which most of the clustered matter is effectively collisionless (dark matter) with the bulk of the energy density of the Universe  behaving like the vacuum energy, $\Lambda$ (dark energy). With only half a dozen parameters, this remarkably simple model is able to explain most of the  different sets of observations spanning over a large range of length scales (for a recent review, see~\cite{Weinberg:2012es}). 

From the theoretical point of view, however, it is also well known that in order to provide a good description of the observed Universe the value of the vacuum energy density, $\rho_\Lambda \simeq 10^{-47}$ Gev$^4$, leads to an unsettled situation in the interface between Cosmology and Particle Physics, since it differs from theoretical expectations by 60-120 orders of magnitude~\cite{Weinberg:1988cp}. Moreover, although the evolution of these two dark components over the cosmic time is significantly different, their current energy densities are of the same order, which  gives rise to the question whether this is only a coincidence or has a more fundamental reason. Such questions are known as the cosmological constant problems~\cite{Weinberg:2000yb}. Thus, given the theoretical uncertainties on the nature and behavior of the dark energy a number of  mechanisms of cosmic acceleration have been investigated, including modifications of gravity on large scales or a possible interaction between the components of the dark sector. 

In particular, interacting models of dark energy and dark matter~\cite{Amendola:1999er,Zimdahl:2001ar, Chimento:2009hj, Ferreira:2012xh, Arevalo:2016epc} are based on the premise that no known symmetry in Nature prevents or suppresses a non-minimal coupling between these components and, therefore, such possibility must be investigated in light of observational data\footnote{A usual critique to these scenarios is that, in the absence of a natural guidance from fundamental physics, one needs to specify a phenomenological interaction term between the dark components  in order to establish a model and study their observational consequences.} (for a recent review, see~\cite{Wang:2016lxa}). In some classes of these coupled models the coincidence problem above mentioned can be largely alleviated when compared with the standard cosmology. This was firstly discussed in reference~\cite{Zimdahl:2001ar}, where the authors investigated asymptotic attractor behaviors for the ratio of the dark matter and dark energy densities. Since then, a number of interacting models with both numerical and analytical solutions have been proposed (see, e.g.~\cite{Jesus:2008xi,Chimento:2009hj,Costa:2009wv,Arevalo:2011hh,Ferreira:2012xh,Marttens:2014yja,Arevalo:2016epc,Marttens:2016cba,Marttens:2017njo,vonMarttens:2018iav} and references therein).

In this paper, we study the observational viability of different classes of interacting scenarios, including both linear and non-linear interaction terms. In order to observationally distinguish between the different models we perform a Bayesian model selection analysis using current data of type Ia supernova (SN Ia)~\cite{Betoule2014,Scolnic:2017caz}, measurements of the baryon acoustic oscillation (BAO) from the 6dFGS \cite{Beutler:2011hx}, SDSS-MGS \cite{Ross:2014qpa}, BOSS-LOWZ \cite{Cuesta:2015mqa}, BOSS-CMASS \cite{Cuesta:2015mqa}, BOSS-DR12 \cite{Alam:2016hwk},  eBOSS \cite{Ata:2017dya,Hou:2018yny} and BOSS-Ly$\alpha$ \cite{Bourboux:2017cbm} surveys, along with BAO measurements obtained from SDSS-DR7~\cite{Alcaniz:2016ryy}, SDSS-DR10~\cite{Carvalho:2015ica}, SDSS-DR11~\cite{Carvalho:2017tuu}, SDSS-DR12Q~\cite{deCarvalho:2017xye}; measurements of the expansion rate from cosmic chronometers~\cite{Moresco:2016mzx} and the estimate of the sound horizon scale at the last scattering reported by the Planck collaboration~\cite{Ade:2015rim}. We find weak and moderate evidence against some of the interacting scenarios studied with respect to $\Lambda$CDM when the full joint analysis is considered. We also discuss how such results can help select viable classes of interacting models.

The paper is organized as follows: in section~\ref{sec:ICS} we present the classes of interacting models studied in this work. In section~\ref{sec:data} we present the datasets considered in the analysis. The Bayesian approach used to evaluate the performance of the different interacting scenarios is described in section~\ref{sec:methodology}. Section~\ref{sec:results} is devoted to the results of the Bayesian comparison analysis. Finally, our results are summarized in section~\ref{sec:finalremarks}.

\section{\label{sec:ICS}Interacting Cosmological Scenarios}
Let us consider a homogeneous, isotropic and flat cosmological scenario described by the Friedmann-Lema\^itre-Robertson-Walker metric (FLRW) and assume that the total energy-momentum tensor for the matter content of the universe is conserved. In the most general case, the total matter density is composed of radiation ($r$), baryons ($b$), dark matter ($dm$) and dark energy ($x$).

Also, let us suppose that the dark components are allowed to interact between each other through a phenomenological interaction term $\Gamma$:
\begin{equation}
    \label{3}
    \rho'_{dm} + \rho_{dm} = -\Gamma \quad \textrm{and} \quad
    \rho'_{x} + \gamma_{x}\rho_{x} = \Gamma \,,
\end{equation}
where prime denotes a convenient derivative with respect to the function of the scale factor $\ln a^3$, $\gamma_x$ is the barotropic index for dark energy and we adopt a pressureless dark matter component. Note that $\Gamma>0$ indicates an energy transfer from dark matter to dark energy and $\Gamma<0$ indicates the opposite. By adding equations~(\ref{3}) and taking $\rho = \rho_{dm} + \rho_x$ we find:
\begin{equation}
    \rho' + \rho_{dm} + \gamma_x\rho_x = 0 \,,
\end{equation}
and, by rearranging the terms, we obtain:
\begin{equation}
    \label{e5}
    \rho_{dm} = -\frac{\gamma_x\rho + \rho'}{1 - \gamma_x} \quad \textrm{and} \quad
    \rho_x = \frac{\rho + \rho'}{1 - \gamma_x} \,.
\end{equation}

If we derivate any of equations~(\ref{e5}) and replace~(\ref{3}) we can write a second order differential equation for a function $\Gamma = \Gamma(\rho, \rho', \rho'')$~\cite{Chimento:2009hj}:
\begin{equation}
    \label{e7}
    \rho'' + (1 + \gamma_x)\rho' + \gamma_x\rho = (1 - \gamma_x)\Gamma \quad \textrm{or} \quad
    \rho(\rho'' + b_1\rho' + b_3\rho) + b_2\rho'^2 = 0 \,,
\end{equation}
where $b_1$, $b_2$, $b_3$ are constants, given in table \ref{T1}, for the classes of interactions that are being considered in this work. The analytical solution of~(\ref{e7}), describing the evolution of the dark sector~\cite{Chimento:2009hj}, takes the following form:
\begin{equation}
    \label{8}
    \rho(a) = 3 H_0^2 (C_1 a^{3\lambda_1} + C_2 a^{3\lambda_2})^{1 / (1 + b_2)} \,,
\end{equation}
where $H_0$ is the Hubble parameter and the constants $C_1$, $C_2$, $\lambda_1$ and $\lambda_2$ are given by:
\begin{eqnarray}
C_1 = (\Omega_{dm0} + \Omega_{x0})^{1 + b_2} - C_2 \,,  \qquad C_2 = \frac{(\Omega_{dm0} + \gamma_x \Omega_{x0})(1 + b_2) + \lambda_1 (\Omega_{dm0} + \Omega_{x0})}{(\Omega_{dm0} + \Omega_{x0})^{-b_2}(\lambda_1 - \lambda_2)} \,,\\ [3ex]
\lambda_1 = \frac{1}{2} \left( -b_1 - \sqrt{b_1^2 - 4b_3 (1 + b_2)} \right) \,, \qquad\qquad \lambda_2 = \frac{1}{2} \left( -b_1 + \sqrt{b_1^2 - 4b_3 (1 + b_2)} \right) \,.
\end{eqnarray}

Throughout this paper, we use the dimensionless density parameters $\Omega_{i0} = \rho_{i0} / 3H_0^2$ with the subscript $0$ denoting their current values. It is worth emphasizing that the evolution of the dark sector is independent of that of radiation and baryons. Finally, the Hubble expansion rate in terms of the cosmological redshift $z$ can be written as:
\begin{equation}
    \label{eq:hubble-expansion}
    H(z) = H_0 \left( \Omega_{r0} (1 + z)^4 + \Omega_{b0} (1 + z)^3 + \left[ C_1 (1 + z)^{-3\lambda_1} + C_2 (1 + z)^{-3\lambda_2} \right]^{\frac{1}{1 + b_2}} \right)^{1/2} \,,
\end{equation}
with $\Omega_{r0} + \Omega_{b0} + \Omega_{dm0} + \Omega_{x0} = 1$. Note that the radiation term includes the contribution of photons, $\Omega_{\gamma0}$, and neutrinos, $\Omega_{\nu0}$.

\begin{table}[!t]
\centering
\caption{\label{T1}Definitions of the parameters $b_1,\ b_2$, $b_3$ for interactions $\Gamma(\rho,\rho',\rho'')$.}
\begin{tabular}{l||c|c|c}\hline\hline
Interaction & $b_1$ & $b_2$ & $b_3$ \\ \hline 
$\Gamma_a=\alpha \rho_{dm}+\beta \rho_{x}$ &$ 1+\gamma_{x}+\alpha-\beta$ & $ 0$ &$\gamma_{x}+\alpha \gamma_{x}-\beta$ \\
[0.5ex] \hline 
$\Gamma_b=\alpha \rho_{dm}^\prime+\beta \rho_{x}^\prime$ &$(1+\gamma_{x}+\alpha\gamma_{x}-\beta)/(1+\alpha-\beta) $&$ 0 $&$\gamma_{x}/(1+\alpha-\beta) $ \\ 
[0.5ex] \hline 
$\Gamma_c=\alpha \rho_{dm}\rho_{x}/\rho$ &$ 1+\gamma_{x} +\alpha(1+\gamma_{x})/(1-\gamma_x)$&$ \alpha/(1-\gamma_x)$&$ \gamma_{x}+ \alpha\gamma_x/(1-\gamma_x)$\\
[0.5ex] \hline 
$\Gamma_d=\alpha \rho_{dm}^2/\rho$ &$ 1 +\gamma_{x} -2\alpha\gamma_{x}/(1-\gamma_x)$&$-\alpha/(1-\gamma_x)$&$\gamma_{x} -\alpha \gamma_{x}^2/(1-\gamma_x) $ \\ [0.5ex]\hline 
$\Gamma_e=\alpha \rho_{x}^2/\rho$ &$1 +\gamma_{x} -2\alpha/(1-\gamma_x)$&$ -\alpha/(1-\gamma_x)$&$\gamma_{x} -\alpha/(1-\gamma_x)$ \\ [0.5ex]
\hline
\hline
\end{tabular}
\end{table}

In reference \cite{Arevalo:2016epc} it was shown that the interacting models in table \ref{T1} can be described in a unified dark sector approach as a variable modified Chaplygin gas \cite{Debnath:2007bw}. The authors point out that these scenarios can also be understood in terms of a varying barotropic index for the dark energy component. Furthermore, most of the interacting models shown in table \ref{T1} have been  proposed in the literature in order to get a scaling solution for the coincidence problem, that is, in the context of these models the universe could approach a stationary stage at which the ratio between dark energy and dark matter densities becomes a constant. In particular, an interacting term of the type $\Gamma\propto\rho_{dm}+\rho_x$ was first introduced in references \cite{Amendola:1999er,Zimdahl:2001ar} from a study of a suitable coupling between a quintessence scalar field and a pressureless cold dark matter field. A generalization of this interaction was considered in reference \cite{Sadjadi:2006qp} in order to overcome the coincidence problem near the transition time in a system that crosses the phantom divide line. %
On the other hand, the interaction $\Gamma\propto\rho'_{dm}+\rho'_{x}$ was first presented in reference \cite{Chimento:2009hj} as a convenient scenario for alleviating the coincidence problem, with an analytical solution for the dark sector. The interaction $\Gamma_c$ was firstly introduced in terms of a non-canonical scaling of the ratio
of the dark matter and dark energy densities as an attempt to solve this same problem in reference \cite{Dalal:2001dt}. This kind of interaction seems to be appealing since it is able to describe the large-scale evolution without instabilities or unphysical features \cite{Li:2013bya}. Finally, interactions of the type $\Gamma_d$ and $\Gamma_e$ were studied in reference \cite{Arevalo:2011hh} in the realm of non-linear interactions alleviating the coincidence problem.

\section{\label{sec:data}Data description}

We focus on background data such as type Ia supernovae through the Joint Light-curve Analysis (JLA) compilation \cite{Betoule2014} and through the Pantheon sample \cite{Scolnic:2017caz}; baryon acoustic oscillation data from 6dFGS \cite{Beutler:2011hx}, SDSS-MGS \cite{Ross:2014qpa}, BOSS-LOWZ \cite{Cuesta:2015mqa}, BOSS-CMASS \cite{Cuesta:2015mqa}, BOSS-DR12 \cite{Alam:2016hwk},  eBOSS \cite{Ata:2017dya,Hou:2018yny} and BOSS-Ly$\alpha$ \cite{Bourboux:2017cbm}; the cosmic chronometers reported in reference~\cite{Moresco:2016mzx} and the angular scale of the sound horizon at the last scattering \cite{Ade:2015rim}. Furthermore, we independently consider BAO measurements obtained by the angular separation between pairs of galaxies \cite{Carvalho:2015ica, Alcaniz:2016ryy, Carvalho:2017tuu, deCarvalho:2017xye}. In what follows, we  briefly present each one of these datasets.

\subsection{BAO data}
\label{BAO}
In table \ref{TableBAOb} of Appendix \ref{sec:3BAO}, the current available measurements of the BAO signal are shown. The isotropic BAO measurements are given in terms of the dimensionless ratio $d_z(z)=D_V(z)/r_d$, where $D_V$ is a combination of the line-of-sight and transverse distance scales defined in reference~\cite{Eisenstein:2005su},
\begin{equation}
D_V(z) = \left( D_M(z)^2 \frac{cz}{H(z)} \right)^{1/3} \,,
\end{equation}
$c$ is the speed of light, $D_M(z) = c\int_0^z \frac{dz}{H(z)}$ is the comoving angular diameter distance and the standard ruler length $r_d$ is usually interpreted as the comoving size of the sound horizon at the drag epoch $r_d=r_s(z_d)$ where
\begin{equation}
\label{rs}
r_s(z) = \int_{z}^{\infty} \frac{c_sdz}{H(z)} \,,
\end{equation}
with $c_s = \frac{c}{\sqrt{3 (1 + \mathcal{R})}}$ being the sound speed of the photon-baryon fluid, $\mathcal{R} = \frac{3\Omega_{b0}}{4\Omega_{\gamma0}(1 + z)}$~\cite{Eisenstein1998} and $z_d$ the redshift at the drag epoch.

As we discuss in Appendix~\ref{sec:3BAO}, BAO measurements obtained through $d_z$ use a fiducial model in order to convert angles into distances. This fact motivates the use of a different set of BAO measurements (claimed to be almost model-independent), which considers only the transversal BAO signal through a geometrical feature such as the angular separation between pairs of galaxies,
\begin{equation}
 \theta_\mathrm{BAO}(z) = \frac{r_d}{D_M(z)} \,.
\end{equation}
As explained in reference \cite{Carvalho:2015ica}, in order to estimate $\theta_{\mathrm{BAO}}$ from $\theta_{\mathrm{FIT}}$ (completely model-independent measurement of the BAO signal obtained from the 2-point angular correlation function) a shift factor is needed, given that the redshift shell-width, $\delta z$, is different from zero. In reference \cite{Carvalho:2015ica} the model-dependence of the shift factor is tested for several cosmologies and the overall conclusion is that the shift factor is almost model-independent, with the difference between $\theta_{\mathrm{BAO}}$ and $\theta_{\mathrm{FIT}}$ being $\lesssim 2\%$ for the considered $\delta z\le 0.02$.

On the other hand, the more recently available anisotropic BAO measurements (BOSS-DR12 and BOSS-Ly$\alpha$ in table \ref{TableBAOb}) consider observables in the transverse direction ($D_M/r_d$) as well as in the radial direction ($D_H/r_d=c/H(z)/r_d$), where both observables are correlated and together they provide complete information about the BAO signature at a given redshift. The corresponding covariance matrix for BOSS-DR12 and BOSS-Ly$\alpha$ are given in references \cite{Alam:2016hwk} and \cite{Bourboux:2017cbm}, respectively.

A relevant issue pointed out in the literature is the lack of informed correlations needed to use all the BAO data combined. In reference~\cite{Ross:2014qpa} the authors argue that SDDS-MGS~\cite{Ross:2014qpa} can be employed in combination with 6dFGS~\cite{Beutler:2011hx} and BOSS-LOWZ~\cite{Cuesta:2015mqa} because the overlapping volumes of the galaxy samples are small enough and consequently the correlations can be considered as negligible. Planck's 2015 collaboration~\cite{Ade:2015xua} uses these BAO data in addition to BOSS-CMASS~\cite{Cuesta:2015mqa} (uncorrelated) to constrain cosmological parameters. In this work we add to the four Planck's 2015 BAO measurements the eBOSS measurement \cite{Ata:2017dya}, which is uncorrelated with the others. From here on we refer to this set of isotropic BAO measurements as BAO3. Additionally we consider a second set of fourteen BAO measurements given by the transverse BAO signal through the angular feature $\theta_{\mathrm{BAO}}$ (see table \ref{BAO2D}). From here on we refer to this set of angular BAO measurement as BAO2. For completeness, we also consider independently a third set of seven isotropic and anisotropic BAO measurements including updated data: 6dFGS~\cite{Beutler:2011hx}, SDSS-MGS~\cite{Ross:2014qpa}, BOSS-DR12 \cite{Alam:2016hwk} replacing BOSS-LOWZ and BOSS-CMASS, eBOSS from reference \cite{Hou:2018yny} and BOSS-Ly$\alpha$ \cite{Bourboux:2017cbm}, we show the results of this last analysis in the Appendix \ref{pantheon}.

Data from the Sloan Digital Sky Survey (SDSS) were employed in obtaining BAO2 measurements (see table~\ref{BAO2D}) as well as some of the BAO3 data points, such as SDSS-MGS, BOSS-LOWZ, BOSS-CMASS, eBOSS. Furthermore, although WiggleZ is an independent survey, in reference~\cite{Beutler:2015tla} the authors pointed out that the volumes considered in the WiggleZ survey and the BOSS-CMASS partially overlap and, as such, they calculated the corresponding correlations. Due to this fact, in this work we assume a conservative approach and do not combine BAO3 and BAO2 measurements.

\begin{table}[!ht]
\centering
\caption{\label{BAO2D} BAO measurements from angular separation of pairs of galaxies.}
\begin{tabular}{lll||lll}
\hline\hline
\multicolumn{1}{c}{$z$}&\multicolumn{1}{c}{$\theta_{\rm{BAO}}(z)\ [^{\circ}]$}&\multicolumn{1}{c||}{Reference}&\multicolumn{1}{c}{$z$}&\multicolumn{1}{c}{$\theta_{\rm{BAO}}(z)\ [^{\circ}]$}&\multicolumn{1}{c}{Reference}\\\hline
0.235&$9.06\pm0.23$&\cite{Alcaniz:2016ryy} SDSS-DR7 &0.550&$4.25\pm0.25$&\cite{Carvalho:2015ica} SDSS-DR10\\
0.365&$6.33\pm0.22$&\cite{Alcaniz:2016ryy} SDSS-DR7 &0.570&$4.59\pm0.36$&\cite{Carvalho:2017tuu} SDSS-DR11\\
0.450&$4.77\pm0.17$&\cite{Carvalho:2015ica} SDSS-DR10 &0.590&$4.39\pm0.33$&\cite{Carvalho:2017tuu} SDSS-DR11\\
0.470&$5.02\pm0.25$&\cite{Carvalho:2015ica} SDSS-DR10 &0.610&$3.85\pm0.31$&\cite{Carvalho:2017tuu} SDSS-DR11\\
0.490&$4.99\pm0.21$&\cite{Carvalho:2015ica} SDSS-DR10&0.630&$3.90\pm0.43$&\cite{Carvalho:2017tuu} SDSS-DR11 \\
0.510&$4.81\pm0.17$&\cite{Carvalho:2015ica} SDSS-DR10&0.650&$3.55\pm0.16$&\cite{Carvalho:2017tuu} SDSS-DR11\\
0.530&$4.29\pm0.30$&\cite{Carvalho:2015ica} SDSS-DR10&2.225&$1.85\pm0.33$&\cite{deCarvalho:2017xye} SDSS-DR12Q \\
\hline\hline
\end{tabular}
\end{table}

In the $\Lambda$CDM scenario the standard ruler $r_d$ coincides with the sound horizon at the drag epoch, which can be determined in a model dependent way from CMB measurements. Nevertheless, in general the two quantities need not coincide \cite{Verde:2016ccp} and some attempts in estimating a model independent low-redshift standard ruler has been made \cite{Heavens:2014rja,Verde:2016ccp}. On the other hand, it has been shown that the tension in $H_0$ could reflect a mismatch between the determination of the standard ruler for the acoustic scale and its standard value \cite{Bernal:2016gxb}. Given this, in this work we consider two different approaches in using the two different BAO datasets, the first one is to calculate the comoving size of the sound horizon \eqref{rs} taking the redshift at the drag epoch given by $z_d = 1059.6$, in accordance with Planck's 2015 results~\cite{Ade:2015xua}. The second approach is to consider the standard ruler for the acoustic scale as a free parameter, where we have chosen as prior the $r_dh$ value reported in reference \cite{Verde:2016ccp} (see table \ref{T4}) instead of Planck's results for the sound horizon since the former was obtained model-independently, just assuming the cosmological principle, a metric theory of gravity and a smooth expansion history without a fiducial cosmology at low redshift. This kind of methodology has been used before in different contexts, e.g., \cite{Tutusaus:2018ulu,Shafer:2015kda}.

\subsection{CMB data}

To perform the full cosmic microwave background analysis building for the entire range of multipoles would require the study of all interacting models at a perturbative level, and also require the adaptation of Boltzmann codes such as CAMB or CLASS to obtain their anisotropy power spectrum. A simpler way to do that is to use the CMB compressed likelihood. 

The compressed likelihood derived from Planck 2015 chains is composed by the set of parameters $\{R,\ell_a,\Omega_{b0}h^2,n_s\}$ \cite{Ade:2015rim}. The shift parameter $R$ is, by construction, very dependent on the matter dominated epoch. This is not evident in Planck's analysis since all models studied have approximately the same behavior in that period. However, when we are dealing with interacting models, the dynamics of the matter dominated epoch is affected, so that constraining $R$ to the values of reference \cite{Ade:2015rim} would include a bias on the analyses, and thereby artificially push the results to resemble the model with no interaction \cite{Ade:2015rim,Sollerman:2009yu}. In order to avoid this, we do not consider the shift parameter in our analysis. Besides, we do not use the spectral index since it does not appear explicitly in our analysis and we fix the physical baryon density to $\Omega_{b0} h^2 = 0.0226$, as reported in \cite{Cooke:2016rky}.

Consequently, this means that the only contribution of CMB data we consider is the angular scale of the sound horizon at the last scattering:
\begin{equation}
    \ell_a = \frac{\pi}{\theta_*} = \frac{\pi D_M(z_*)}{r_s(z_*)} \,,
\end{equation}
where the comoving size of the sound horizon is evaluated at the redshift of last scattering $z_*$. We compare the value obtained in our study to the one reported by the Planck collaboration in 2015, $\ell_a=301.63\pm 0.15$ \cite{Ade:2015rim}.
In order to elucidate if there is a noticeable change in our results in fixing or estimating $z_*$ we use two different methods, the first one considers $z_* = 1089.9$, according with Planck's 2015 results \cite{Ade:2015xua}. The second one contemplates the computation of the redshift at drag epoch from the following expression \cite{Hu:1995en},
\begin{eqnarray}
\label{zdec}
z_{*}=1048(1 + 0.00124(\Omega_{b0} h^2)^{-0.738})(1+ g_1(\Omega_{m0} h^2)^{g_2}),
\end{eqnarray}
where we have considered $\Omega_{m0}=\Omega_{b0}+\Omega_{dm0}$ and 
\begin{eqnarray}
g_1=\frac{0.0783(\Omega_{b0} h^2)^{-0.238}}{1 + 39.5(\Omega_{b0} h^2)^{0.763}} \quad     \textrm{and}\quad
g_2=\frac{0.560}{1+ 21.1(\Omega_{b0} h^2)^{1.81}}.
\end{eqnarray}
This expression has been used before in studying interacting scenarios (see e.g. \cite{Arevalo:2016epc,Santos:2016sti}). Notice that \eqref{zdec} is weakly dependent on the dark matter contribution, the difference with Planck's 2015 result for $z_*$ is lower than 1$\%$ for $0.1<\Omega_{dm0}<0.4$ and $0.6<h<0.8$.

\subsection{Cosmic chronometers}

All data from cosmic chronometers used here were obtained through the differential age method (see table \ref{tab:h_clocks}). We were careful to use only these measurements, and not to include those obtained with BAO so as not to double count information in the joint analyses.

The procedure consists of taking the relative age of passively evolving galaxies, with respect to the redshift, as suggested by reference~\cite{jimenez02}. Most of the values are obtained from the {BC03} catalogue \cite{bc03}, but values from older releases were also used \cite{Simon:2004tf}. With the ratio between the differential ages, $dt$ and the respective difference in redshift, $dz$, obtaining $H(z)$ becomes a simple task:
\begin{equation}
H(z) = - \frac{1}{1 + z} \frac{dz}{dt} \,.
\end{equation}
In our analysis, the theoretical value of $H(z)$ is given by equation~(\ref{eq:hubble-expansion}).
Note that we only use data of redshift up to $< 1.2$. This is motivated by a discussion in reference~\cite{Verde:2014qea}, where the authors argue that the expansion history data of the universe might not be necessarily smooth outside $0.1 < z < 1.2$. Likewise, the authors of reference~\cite{Moresco:2012jh} have also shown that outside this range, the model of synthesis of stellar population adopted to derive the galaxy ages becomes relevant.
\begin{table}[ht!]
\centering
\caption{Estimated values of $H(z)$ obtained using the differential age method.}
\begin{tabular}{lll||lll||lll}
\hline
\hline
\multicolumn{1}{c}{$z$} & \multicolumn{1}{c}{$H$ [km/s/Mpc]} & \multicolumn{1}{c||}{Ref.} & \multicolumn{1}{c}{$z$} & \multicolumn{1}{c}{$H$ [km/s/Mpc]} & \multicolumn{1}{c||}{Ref.} & \multicolumn{1}{c}{$z$} & \multicolumn{1}{c}{$H$ [km/s/Mpc]} & \multicolumn{1}{c}{Ref.}
\\ \hline
$0.07   $ & $ 69    \pm 19.6 $ & \cite{Zhang:2012mp} &  $0.28   $ & $ 88.8  \pm 36.6 $ & \cite{Zhang:2012mp} & $0.48    $ & $ 97    \pm 62   $ & \cite{stern10}         \\ [1ex]
$0.09   $ & $ 69    \pm 12   $ & \cite{Simon:2004tf} &  $0.352  $ & $ 83    \pm 14   $ & \cite{Moresco:2012jh}   & $0.593  $ & $ 104   \pm 13   $ & \cite{Moresco:2012jh}       \\ [1ex]
$0.12   $ & $ 68.6  \pm 26.2 $ & \cite{Zhang:2012mp} &  $0.3802 $ & $ 83    \pm 13.5 $ & \cite{Moresco:2016mzx} & $0.68     $ & $ 92    \pm 8    $ & \cite{Moresco:2012jh}  \\ [1ex]
$0.17   $ & $ 83    \pm 8    $ & \cite{Simon:2004tf} &  $0.4    $ & $ 95    \pm 17   $ & \cite{Simon:2004tf} & $0.781  $ & $ 105   \pm 12   $ & \cite{Moresco:2012jh}       \\ [1ex]
$0.179  $ & $ 75    \pm 4    $ & \cite{Moresco:2012jh}    & $0.4004 $ & $ 77    \pm 10.2 $ & \cite{Moresco:2016mzx} & $0.875  $ & $ 125   \pm 17   $ & \cite{Moresco:2012jh}        \\ [1ex]
$0.199  $ & $ 75    \pm 5    $ & \cite{Moresco:2012jh}   &  $0.4247 $ & $ 87.1  \pm 11.2 $ & \cite{Moresco:2016mzx} & $0.88     $ & $ 90    \pm 40   $ & \cite{stern10}\\ [1ex]
$0.20   $ & $ 72.9  \pm 29.6 $ & \cite{Zhang:2012mp} &  $0.4497$ & $ 92.8  \pm 12.9 $  & \cite{Moresco:2016mzx} & $0.9      $ & $ 117   \pm 23   $ & \cite{Simon:2004tf}\\ [1ex]
$0.27   $ & $ 77    \pm 14   $ & \cite{Simon:2004tf} &  $0.4783 $ & $ 80.9  \pm 9    $ & \cite{Moresco:2016mzx} & $1.037  $ & $ 154   \pm 20   $ & \cite{Moresco:2012jh} \\ [1ex]
\hline
\hline
\end{tabular}
\label{tab:h_clocks}
\end{table}

Notice that the cosmic chronometer is  the  only method providing cosmology-independent, direct measurements of the expansion history of the universe~\cite{Verde:2014qea}.

\subsection{Supernovae Ia}

The best probe of the expansion history of the Universe on large-scales (up to $z \lesssim 2$) is provided nowadays by observations of type Ia supernovae.
The main reason for this is that SNe Ia are examples of "standardisable candles", due to the fact that their absolute magnitudes can be approximated by using light-curve templates to extract their stretch and color parameters.
We use the JLA sample which contains a set of 740 spectroscopically confirmed SNe Ia \cite{Betoule2014} composed by several low-redshift ($z < 0.1$) samples, the full three-year SDSS-II supernova survey sample~\cite{Sako:2014qmj} in the interval $0.05 < z < 0.4$, the three-years data of the SNLS survey~\cite{Conley2011, Guy2010} up to redshift $z < 1$ and some high-redshift Hubble Space Telescope SNe~\cite{Riess2007} with redshift $0.216 < z < 1.755$.

The predicted apparent magnitude of a SN Ia can be obtained from its light curve parameters through the linear relation:
\begin{equation}
    \label{eq:data.SN.apparent_magnitude}
    m(z, \Theta) = \mu(z, \Theta) + M_B - \aJLA \times x_1 + \bJLA \times c \,,
\end{equation}
where $\Theta$ represents the set of parameters of the model, $x_1$ is the time stretching of the light curve and $c$ is the supernova color at its maximum brightness. In the expression above, $\mu(z, \Theta)$ is the theoretical distance modulus, given by:
\begin{equation}
    \label{eq:data.SN.distance_modulus}
    \mu(z, \Theta) = 5 \log \frac{d_\mathrm{L}(z, \Theta)}{10\,\mathrm{pc}} \,,
\end{equation}
where $d_\mathrm{L}(z, \Theta)$ is the luminosity distance:
\begin{equation}
    \label{eq:data.SN.luminosity_distance}
    d_\mathrm{L}(z) = (1 + z) \int_0^z \frac{\mathrm{d}z'}{E(z')} \,,
\end{equation}
and $E(z) \equiv H(z) / H_0$.

In equation~(\ref{eq:data.SN.apparent_magnitude}), the light-curve parameters $x_1$ and $c$ have different values for each supernova and are derived directly from the light-curves.
However, the nuisance parameters $M_B$, $\aJLA$ and $\bJLA$ are assumed to be constant for all the supernovae but differ for each cosmological model.
Additionally, since the properties of the host galaxy can generate some effects on the intrinsic brightness of the SNe Ia, we follow reference~\cite{Betoule2014} and model the relation between $M_B$ and the host galaxy stellar mass, $M_\mathrm{host}$, by assuming $M_B \rightarrow M_B + \Delta_M$ if $\log_{10} M_\mathrm{host} > 10$.
Thus, the nuisance parameters corresponding to the measurements  are $\aJLA$, $\bJLA$, $M_B$ and $\Delta_M$.

The analyses involving the JLA dataset were carried out by comparing the predicted magnitude $m(z, \Theta)$ from equation~(\ref{eq:data.SN.apparent_magnitude}) against the observed ones of the JLA sample (table F.3 of reference~\cite{Betoule2014}), which are denoted by $m_B(z)$ and represent the observed peak magnitude in rest-frame $B$ band.
The Monte Carlo analyses for the JLA SNe Ia sample were performed by assuming a multivariate Gaussian likelihood of the form
\begin{equation}
    \label{eq:data.SN.likelihood}
    \mathcal{L}_\mathrm{JLA}(D \mid \Theta) = \exp \left[-\chi_\mathrm{JLA}^2(D \mid \Theta) \,/\, 2 \right] \,,
\end{equation}
with
\begin{equation}
    \label{eq:data.SN.chi2}
    \chi^2_\mathrm{JLA}(\Theta) = \left[ \mathbf{m}_B - \mathbf{m}(\Theta) \right]^T C^{-1} \left[ \mathbf{m}_B - \mathbf{m}(\Theta) \right] \,,
\end{equation}
where $C$ corresponds to the covariance matrix of the $m_B$ measurements, estimated accounting for various statistical and systematic uncertainties (we refer the reader to reference~\cite{Betoule2014} for more information about these uncertainties).

Since the Pantheon supernovae sample \cite{Scolnic:2017caz} came to light during the development of this work, for completeness we include in Appendix \ref{pantheon} a brief description of this dataset and the results when replacing the JLA compilation with the Pantheon compilation.

\subsection{Joint analysis}
By using the same methodology as in the case of the JLA SNe compilation, we consider a multivariate Gaussian likelihood for BAO2, BAO3, CC and CMB data. The chi-square function for the measurement of a generic function $f$ is defined as follows:
\begin{eqnarray}
\label{chi}
\chi^2_f(\Theta)=\sum_i\left(\frac{f(z_i)-f(z_i,\Theta)}{\sigma_{f_i}}\right)^2,
\end{eqnarray}
where $f(z_i)$ represents the measured value for $f$ at redshift $z_i$, whereas $f(z_i,\Theta)$ is computed assuming a model with parameters $\Theta$.  The function $f$ stands for the functions $\theta_{\rm{BAO}}(z)$, $d_z(z)$, $H(z)$ and $\ell_a(z_*)$ for BAO2, BAO3, CC and CMB data, respectively. The sum in equation~(\ref{chi}) runs over the data in table \ref{BAO2D} for BAO2 and table \ref{tab:h_clocks} for CC. For BAO3 we consider 6dFGS, SDSS-MGS, BOSS-LOWZ, BOSS-CMASS and eBOSS as described in section \ref{BAO}. Independently, in the Appendix \ref{pantheon} we consider the following set of updated BAO measurements: 6dFGS, SDSS-MGS, BOSS-DR12, eBOSS and BOSS-Ly$\alpha$. For CMB, we take just a single data at $z_*$. In the case of the joint analysis, the total likelihood is obtained as the product of individual likelihoods associated to each data as in equation~(\ref{eq:data.SN.likelihood}) by using the chi-square definition in equation~(\ref{chi}). For example, the full joint analysis considering BAO2 data is given by: $\mathcal{L}_{\rm{joint}}=\mathcal{L}_{m(z)}\times\mathcal{L}_{\theta_{\rm{BAO}}(z)}\times\mathcal{L}_{H(z)}\times\mathcal{L}_{\ell_a(z_*)}$.

\section{\label{sec:methodology}Methodology: Bayesian model selection}

The Bayesian inference method constitutes a robust statistical technique for parameter estimation and model selection, and over the last years has been widely used in the study of cosmological scenarios \cite{Santos:2016sti,Heavens:2017hkr,SantosdaCosta:2017ctv,Andrade:2017iam,Ferreira:2017yby}. Bayesian inference is based on the Bayes' theorem, which updates our knowledge of a given model (or hypothesis) in light of new available data (or information). Mathematically, the Bayes' theorem gives us the posterior probability $P$ for a set of parameters $\Theta$, given the data $\mathcal{D}$, described by a model $\mathcal{M}$,
\begin{eqnarray}
\label{posterior}
P(\Theta\vert \mathcal{D},\mathcal{M})=\frac{\mathcal{L}(\mathcal{D}\vert\Theta,\mathcal{M})\ \mathcal{P}(\Theta\vert\mathcal{M})}{\mathcal{E}(\mathcal{D}\vert\mathcal{M})},
\end{eqnarray}
{where $\mathcal{L}$, $\mathcal{P}$ and $\mathcal{E}$ stand for the likelihood, prior and evidence, respectively.}

The evidence in equation~(\ref{posterior}) constitutes just a normalization constant in the Bayesian parameter estimation approach, however, it becomes a key ingredient in the Bayesian model comparison approach. In order to compare the performance of different models given a set of data, we use the Bayes' factor defined in terms of the evidence of models $\mathcal{M}_i$ and $\mathcal{M}_j$ as:
\begin{eqnarray}
\label{Bayes}
B_{ij}=\mathcal{E}_i/\mathcal{E}_j,
\end{eqnarray}
{where the evidence corresponds to the average value of the likelihood over the entire model parameter space allowed, before we observe the new data \cite{Liddle:2007fy}, that is:}
\begin{eqnarray}
\mathcal{E}(\mathcal{D}\vert\mathcal{M})=\int \mathcal{L}(\mathcal{D}\vert\Theta,\mathcal{M})\ \mathcal{P}(\Theta\vert\mathcal{M})\ d \Theta.
\end{eqnarray}
{If the models $\mathcal{M}_i$ and $\mathcal{M}_j$ have the same prior probability, then the Bayes' factor gives the posterior odds of the two models.}

Monte Carlo sampling techniques are widely applied nowadays to construct the posterior distribution in equation~(\ref{posterior}), since it is very difficult to compute the posterior numerically (see references~\cite{Lewis2002, Mukherjee2006} for applications of some Monte Carlo algorithms in cosmology).
In this sense, we performed the analyses involving the data described in section~\ref{sec:data} by applying the nested sampling (NS) Monte Carlo algorithm~\cite{Skilling2004}, which is well known for its efficiency in the evidence computation since it is designed to directly estimate the relation between the likelihood function and the prior mass, thus obtaining the evidence (and its uncertainty) immediately by summation, while also computing the samples from the posterior distribution as an optional by-product.
To compute the evidence values and generate the posterior distributions we used the \MN\footnote{\url{https://ccpforge.cse.rl.ac.uk/gf/project/multinest}.}~\cite{Feroz2008, Feroz2009, Feroz2013} algorithm, requiring a global log-evidence tolerance of 0.01 as a convergence criterion and working with a set of 1000 live points to improve the accuracy in the estimate of the evidence.
With this number of live points, the number of samples for all posterior distributions was of order $\mathcal{O}(10^4)$.

The Jeffreys' scale \cite{Jeffreys61} gives us an empirical measure for interpreting the strength of the evidence in comparing two competing models. In order to perform model comparison in this work we use a conservative version of the Jeffreys' scale defined in reference \cite{Trotta:2008qt} (see table \ref{jeffreys}). Usually for $\vert B_{ij}\vert<1$ the evidence in favor/against model $\mathcal{M}_i$ relative to model $\mathcal{M}_j$ is interpreted as inconclusive. On the other hand, the thresholds shown in table \ref{jeffreys} for weak,  moderate and strong evidences in favor/against the tested model correspond to posterior odds of about 3:1, 12:1 and 150:1, respectively \cite{Trotta:2008qt}.  Here, we take $\Lambda$CDM as the reference model $\mathcal{M}_j$, as such, the subscripts in the Bayes' factor (\ref{Bayes}) will be omitted hereafter. Note that from now on, $\ln B<-1$ means support in favor of the $\Lambda$CDM model. 
\begin{table}[ht!]\centering
\caption{\label{jeffreys} The Jeffreys' scale, empirical measure for interpreting the evidence in comparing two models $\mathcal{M}_i$ and $\mathcal{M}_j$ as presented in reference~\cite{Trotta:2008qt}. The left column indicates the threshold for the logarithm of the Bayes factor and the right column the interpretation for the strength of the evidence above the corresponding threshold.}
    \begin{tabular}{c c}        \hline\hline
        $\abs{\ln B_{ij}}$ & Interpretation\\
        \midrule
        $<1$               & inconclusive\\
        $1$                & weak\\
        $2.5$              & moderate\\
        $5$                & strong\\        \hline\hline
    \end{tabular}
\end{table}

In this work, we consider two different approaches in using the data, where the goal is to elucidate if there is a noticeable impact in our final results by considering different numbers of free parameters and different kinds of priors for some specific parameters. In our first approach, we use the Planck's values \cite{Ade:2015xua} for the redshift at the drag epoch and last scattering, $z_d = 1059.60$ and $z_* = 1089.90$, respectively. The priors considered here are shown in table \ref{T4}. We have  chosen a uniform prior for the parameters appearing in all the studied models such as $\Omega_{dm0}$, $h$, $\alpha_{JLA}$, $\beta_{JLA}$, $M_B$ and $\Delta_M$, and a Gaussian prior for the parameters defining only some of the models, i.e., $\gamma_x$ and the interacting parameters $\alpha$ and $\beta$. 
For the parameter $\Omega_{dm0}$ we choose a conservative uniform prior between 0 and 1, and for the dimensionless Hubble parameter $h$ we adopt a range 10 times wider than the 1$\sigma$ value reported by Riess et al. in reference~\cite{riess16}; we emphasize that Riess' result is considered only through this prior and not used  as an independent data in the analysis. For the JLA parameters, $\alpha_{JLA}$, $\beta_{JLA}$, $M_B$ and $\Delta_M$, we use a range 20 times wider than the $1\sigma$ values reported by Betoule et al. in reference~\cite{Betoule2014}. The prior for the parameter $\gamma_x$ corresponds to the 1$\sigma$ value informed by Planck \cite{Ade:2015xua} and the prior for the interacting parameters is the same for all the studied models, of order $10^{-3}$ with a negative mean (see reference~\cite{Arevalo:2016epc}), in such a way that most of the models in table \ref{T1}, $\Gamma_a$, $\Gamma_c-\Gamma_e$, favor a transfer from dark energy to dark matter provided that $\rho_{dm}$ and $\rho_x$ are positive defined. This kind of transfer is expected for some interacting models based on thermodynamical arguments \cite{Pavon:2007gt}. 

In the second approach we change the prior for the $h$ parameter to a Gaussian prior (or equivalently we consider the value reported by Riess et al. in reference~\cite{riess16} as a data), we compute $z_*$ from \eqref{zdec} instead of fixing it and we consider $r_dh$ a free parameter with a Gaussian prior as in reference \cite{Verde:2016ccp}.

In a nutshell, we study the following scenarios as two independent sets of   priors (see table \ref{T4}) and definitions,
\begin{eqnarray}
&\ \textrm{Gaussian prior for } \{\gamma_x,\alpha,\beta\};\ r_d\ \textrm{obtained from \eqref{rs}};\ z_*,z_d\ \textrm{fixed}\label{sc1}\\
&\ \textrm{Gaussian prior for } \{h,\gamma_x,\alpha,\beta,r_d h\};\ z_*\ \textrm{obtained from \eqref{zdec}}\label{sc2}
\end{eqnarray}
Both scenarios consider uniform priors for the remaining parameters defined in table \ref{T4}.

Given that we expect interacting models not to affect the physics of the primordial universe but only to modify the evolution of the dark sector recently, we fix the following parameters: $\Omega_{b0} h^2 = 0.0226$ \cite{Cooke:2016rky}, $\Omega_{\gamma0} h^2 = 2.469 \times 10^{-5}$ \cite{Komatsu:2010fb}, $\Omega_{r0} = \Omega_{\gamma0} \left( 1 + \frac{7}{8} \left( \frac{4}{11} \right)^{\frac{4}{3}} N_\mathrm{eff} \right)$, $N_{\mathrm{eff}} = 3.046$ \cite{Mangano:2005cc}. 
\begin{table}[!ht]
\centering
\caption{ Priors on the free parameters of the studied models. For a Gaussian prior we inform $(\mu,\sigma^2)$ and for a Uniform prior $(a,b)$ represent $a\le x\le b$.}
\begin{tabular}{c |l |l |c}
\hline\hline
Parameter & Status & Prior & Ref. \\\hline
$\Omega_{dm0}$&Global Parameter&Uniform: $(0,1)$& -\\
$h$ &Global Parameter&Uniform: $(0.5584,0.9064)$&\cite{riess16}\\
 &&Gaussian: $(0.7324,0.0174)$&\cite{riess16}\\
$\gamma_x$ &Variable state parameter &Gaussian: $(-0.006,0.002)$& \cite{Ade:2015xua}\\
$\alpha$ &Interacting models&Gaussian: $(-0.001,0.01)$&\cite{Arevalo:2016epc}\\
$\beta$  &Interacting models&Gaussian: $(-0.001,0.01)$&\cite{Arevalo:2016epc}\\
$\alpha_{JLA}$ &Global, JLA parameter&Uniform: $(0.021,0.261)$&\cite{Betoule2014}\\
$\beta_{JLA}$ &Global, JLA parameter&Uniform: $(1.601,4.601)$&\cite{Betoule2014}\\
$M_B$ &Global, JLA parameter&Uniform: $(-19.45,-18.65)$&\cite{Betoule2014}\\
$\Delta_M$ &Global, JLA parameter&Uniform: $(-0.53,0.39)$& \cite{Betoule2014}\\
$r_d h$&Global Parameter&Gaussian: $(102.3,1.6)$&\cite{Verde:2016ccp}\\
\hline\hline
\end{tabular}
\label{T4}
\end{table}

\section{\label{sec:results}Analysis and Results}

We perform a Bayesian comparison analysis of the interacting models presented in table~\ref{T1} in terms of the strength of the evidence according to Jeffreys' scale (see table \ref{jeffreys}). In this study we used the priors shown in table \ref{T4} and considered different combinations of background data such as, type Ia supernovae, cosmic chronometers, baryon acoustic oscillations and cosmic microwave background. Our main results are summarized in tables~\ref{Tf1}--\ref{full_nJB3}. We labeled different realizations of models in table~\ref{T1} with numerical subscripts as follows: 0, 1, 2, 3, 4 meaning $\gamma_x = 0$, $\beta = 0$, $\alpha = 0$, $\alpha = \beta$, $\alpha \neq \beta$, respectively. We do not include the analysis for the interacting model $\Gamma_{b02}$ because it reduces to the $\Lambda$CDM scenario.

In tables \ref{Tf1}--\ref{full3}, the results were obtained using the priors given by \eqref{sc1}, while tables \ref{full_nJB2}--\ref{full_nJB3} the priors used were those of \eqref{sc2}. In tables \ref{full}--\ref{full_nJB3} we have not reported the nuisance parameters because the variation in the best fit estimation among different models is negligible (see Appendix \ref{nuisance2}).

In table \ref{Tf1} we observe that by considering JLA, BAO2 or CC by themselves or even the joint analysis with BAO2 + CC, we get inconclusive evidence for the interacting models and for the $\omega$CDM model when compared to $\Lambda$CDM. We remark that the BAO2 and CC data that were used are almost model-independent and model-independent, respectively, and these alone or jointly are not able to rule out any of the models considered in this work compared to $\Lambda$CDM. We also observe that by analyzing BAO2 + CC + CMB we find weak or moderate evidence disfavoring some of the interacting models.

In table \ref{Tf2} we show the logarithm of the Bayesian evidence and the interpretation of the strength of the evidence for the analysis with BAO3, BAO3 + CC and BAO3 + CC + CMB. We note that some of the interacting models are disfavored with a weak evidence in the studies with BAO3 and BAO3 + CC, and these evidences become moderate (in most of the cases) when we add CMB data to the analysis. Moreover, we notice that the models presenting inconclusive evidence in this analysis are the same as in the study of BAO2 + CC + CMB, which seems to indicate that the addition of the CMB data to the analysis, even when it is a single data point, contributes significantly to the evidence disfavoring most of the interacting models. A different example of the importance in considering CMB measurements is studied in reference \cite{Tutusaus:2016orn}, where the authors show that a power law scenario is supported by SNe Ia alone or combined with BAO data, nevertheless the addition of a single CMB measurement to the joint analysis rules out this scenario.

In tables \ref{full}--\ref{full3} the results for the full joint analysis, including SNe Ia from JLA compilation, are shown considering independently the measurements of BAO2 and BAO3, respectively. We present the best fit parameters for the studied models, along with the logarithm of the Bayesian evidence, the logarithm of the Bayes' factor and the interpretation for the strength of the evidence. It is interesting to note that, for the results in table \ref{full} as well as the results in table \ref{full3}, the evidence remains inconclusive for the models $\omega$CDM, $\Gamma_{a02}$, $\Gamma_{c0}$, $\Gamma_{e0}$, $\Gamma_{a2}$, $\Gamma_{b2}$, $\Gamma_{c}$ and $\Gamma_{e}$ when compared to $\Lambda$CDM. These models are the same as those presenting inconclusive evidence in the analyses BAO2 + CC + CMB and BAO3 + CC + CMB. Observe that the inconclusive models correspond to $\Gamma\propto\rho_x$ or $\Gamma\propto\rho_x'$ and the energy transfer for these models turns out to be from dark energy to dark matter. In this context, in the recent work \cite{vonMarttens:2018iav} interacting scenarios $\Gamma_{a10},\ \Gamma_{a20},\ \Gamma_{c0},\ \Gamma_{d0}$ and $\Gamma_{e0}$ were studied, performing a joint analysis with the full CMB temperature anisotropy spectrum, JLA and BAO3. The authors found, using theoretical arguments and observational results, that while models with $\Gamma\propto\rho_m$ are virtually discarded, there is still room for models with $\Gamma\propto\rho_x$. 

From tables \ref{full}--\ref{full3} we notice that the moderated evidence found in the analysis BAO2 + CC + CMB becomes weak when we consider the full analysis including JLA. On the other hand, the weak evidences obtained in the analysis with JLA + BAO2 + CC + CMB become all moderate when we take BAO3, instead of BAO2 in the full joint analysis. This result is expected since BAO3 data implicitly seem to favor the $\Lambda$CDM scenario. 
It is worth pointing out here that the value of the Hubble parameter turns out to be higher for all the studied scenarios when we consider BAO2 instead of BAO3 in the full joint analysis (see tables \ref{full} and \ref{full3}), this seems to indicate that the BAO data play an important role in the $H_0$ tension \cite{Benetti:2017juy}, we also observe that the values of the fitted barotropic index and the interacting parameters are approximately one order of magnitude lower in the full joint analysis including BAO3 compared to the analysis with BAO2, which reinforce the indication for a preference of $\Lambda$CDM in BAO3 data.

In tables \ref{full_nJB2}--\ref{full_nJB3} the analogous results, to those in tables \ref{full}--\ref{full3}, for the full joint analysis with the JLA compilation  are shown considering independently the measurements of BAO2 and BAO3, respectively. The differences between tables \ref{full}--\ref{full_nJB2} and \ref{full3}--\ref{full_nJB3} rely on considering scenario \eqref{sc1} instead of scenario \eqref{sc2} for the priors. We observe that in tables  \ref{full_nJB2}--\ref{full_nJB3}, as in tables \ref{full}--\ref{full3}, the evidence remains inconclusive for the models $\omega$CDM, $\Gamma_{a02}$, $\Gamma_{c0}$, $\Gamma_{e0}$, $\Gamma_{a2}$, $\Gamma_{b2}$, $\Gamma_{c}$ and $\Gamma_{e}$ when compared to $\Lambda$CDM. However, from tables \ref{full_nJB2}--\ref{full_nJB3} it is not so clear that BAO3 data favor the $\Lambda$CDM scenario, a similar behavior is observed in tables \ref{full_PB2}--\ref{full_PB3} in Appendix \ref{pantheon} where, instead of the JLA, we use the most up to date SNe Ia sample (Pantheon).

In tables \ref{full_nJB2} and \ref{full_nJB3} we notice that in the best fit estimation for the $h$ parameter, the error is in general of order 1-2$\%$ and the dispersion (comparing higher to lower estimation for all the models) is always lower than 1$\%$. The $r_d$ parameter presents a similar behavior, the error is around 1$\%$ and the dispersion around $0.5\%$. Nevertheless, when comparing the best fit for $h$ between tables  \ref{full_nJB2} and \ref{full_nJB3} the difference is around 1$\%$, while for $r_d$ the best fit is close to $3\%$. From this, we conclude that in comparing $\Lambda$CDM with interacting models there is no prominent difference in the best fit estimation for the parameters $h$ or $r_d$ and while there is no distinctive variation in the $h$ estimation in using the full analysis with BAO2 or BAO3, the standard ruler $r_d$ becomes closer to the prior (and also closer to the Planck's 2015 estimation \cite{Ade:2015xua}) when using BAO3 data. An analogous behavior is observed from tables \ref{full_PB2}--\ref{full_PB3} in the Appendix \ref{pantheon}. Also, there is no noticeable difference in the order of magnitude in estimating parameters $\alpha$, $\beta$ or $\gamma_x$ (see tables \ref{full_nJB2}--\ref{full_nJB3} and \ref{full_PB2}--\ref{full_PB3}).

For completeness, in table \ref{full_newBAO} of Appendix B we have included the full joint analysis with the set of priors \eqref{sc2} and an updated set of BAO measurements, including data from: 6dFGS \cite{Beutler:2011hx}, SDSS-MGS \cite{Ross:2014qpa}, BOSS-DR12 \cite{Alam:2016hwk},  eBOSS \cite{Hou:2018yny} and BOSS-Ly$\alpha$ \cite{Bourboux:2017cbm}. The differences in parameter estimation between tables \ref{full_PB3} and \ref{full_newBAO} are negligible, the major change is presented in the interpretation of the strength of the evidence, in those cases where the evidence is weak in table \ref{full_PB3} it becomes moderate in table \ref{full_newBAO}, in both cases the evidence is supporting $\Lambda$CDM.

In reference~\cite{Arevalo:2016epc} some of the models studied in this work were analyzed in a joint analysis considering Union2.1 + BAO3 + H(z) + CMB, and a comparison among the  models was performed using the Akaike and Bayesian information criteria (AIC and BIC, respectively). At this point, it is worth mentioning that, differently from these methods, the Bayesian model comparison applied in this analysis selects the best-fit model by comparing the compromise between quality of the fit and predictability, and by evaluating if the extra degrees of freedom of a given model are indeed required by the data, preferring the model that describes the data well over a large fraction of their prior volume. The results of~\cite{Arevalo:2016epc} in terms of BIC indicate that the interacting models have "strong evidence against" when compared to $\Lambda$CDM and this strength of evidence changes to "evidence against" (or moderate) when compared to $\Lambda$CDM but using binned JLA data instead of Union2.1. Besides, the BAO measurements used included  WiggleZ data, but the full JLA dataset was not considered.
 
The authors also compared the interacting models with the $\omega$CDM model obtaining "not enough evidence against" or inconclusive evidence for the interacting models with three free parameters analyzed ($\Gamma_{a01}$, $\Gamma_{a02}$, $\Gamma_{a03}$, $\Gamma_{b01}$, $\Gamma_{c0}$, $\Gamma_{d0}$ and $\Gamma_{e0}$ in this work). In our study, the evidence interpretation remains the same in most of the models when we compare them to $\omega$CDM instead of $\Lambda$CDM.  
Finally, the ordering of the evidence in terms of the number of free parameters for the interacting models reported in reference~\cite{Arevalo:2016epc} with the BIC approach is not observed in our work using a full Bayesian comparison approach.

In reference~\cite{Santos:2016sti} the model $\Gamma_{a01}$ was analyzed in a model comparison approach with JLA data and a different set of BAO3 measurements. The authors found moderate evidence in favor of this model compared to $\Lambda$CDM. Instead, in this work we found moderate evidence disfavoring the $\Gamma_{a01}$ model. The main differences between our datasets and the ones used in reference~\cite{Santos:2016sti} are that (i) we are considering CMB data and high redshift values of BAO2 and BAO3 data that have become available only recently. Furthermore, the authors of reference~\cite{Santos:2016sti} used WiggleZ data, which is not considered in the set of BAO3 data used in this work. Besides, the authors use the Eisenstein's approximation for the redshift at the drag epoch, whereas here we use the Planck's value in scenario \eqref{sc1} or $r_dh$ as a free parameter in scenario \eqref{sc2}.

\begin{figure}[ht!]
\centering
\includegraphics[scale=0.7]{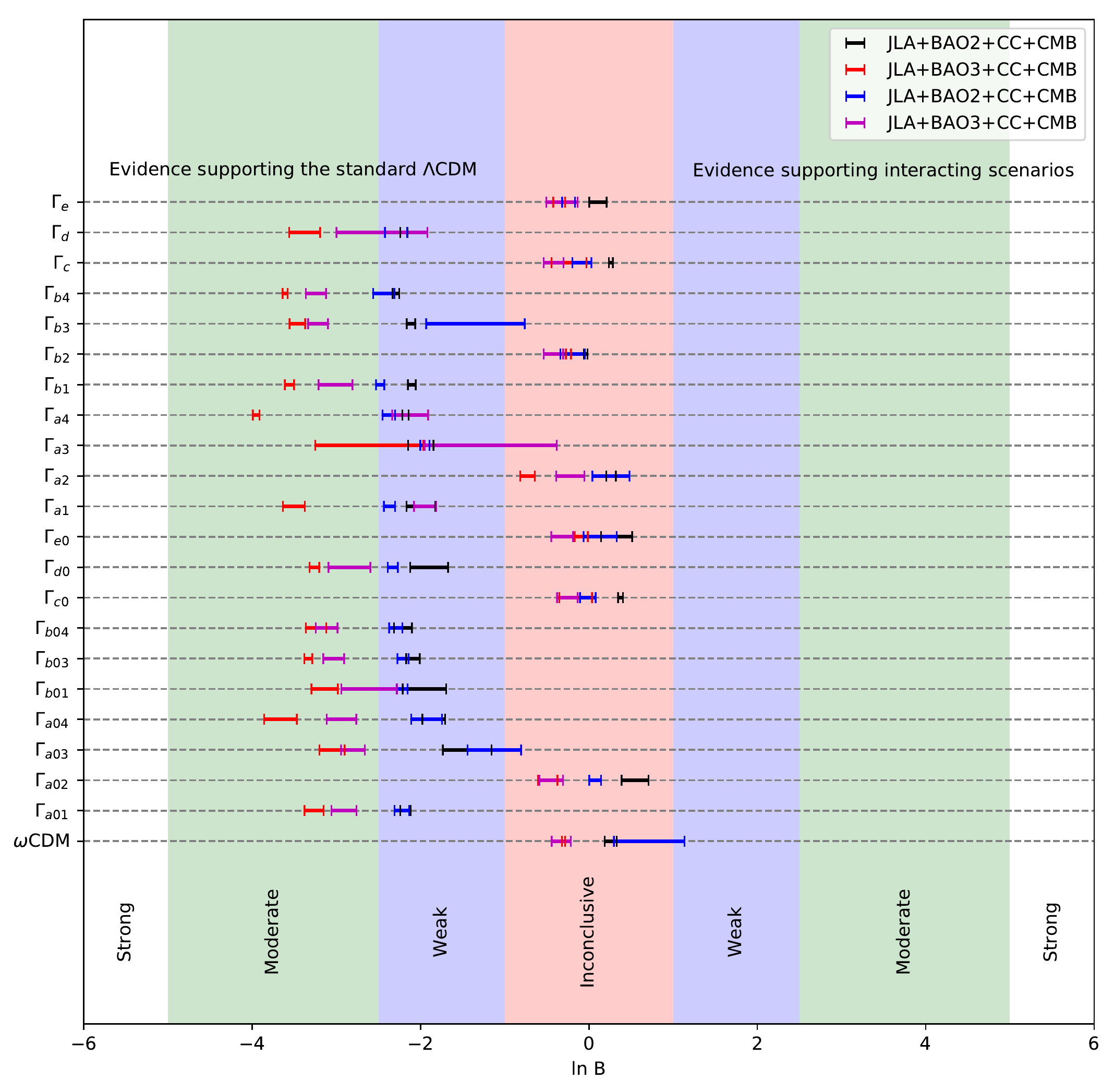}
\caption{\label{comparison} Summary of Bayesian comparison between interacting models and $\Lambda$CDM. We show the classification in terms of the Jeffreys' scale, where $\ln B<-1$ favors the $\Lambda$CDM scenario. The colored intervals represent the 1$\sigma$ region in the estimation of the Bayes' factor. Intervals in black and red consider as priors scenario \eqref{sc1} and those in blue and magenta consider as prior scenario \eqref{sc2}.}
\end{figure}

\begin{figure}[ht!]
\centering
\includegraphics[scale=0.45]{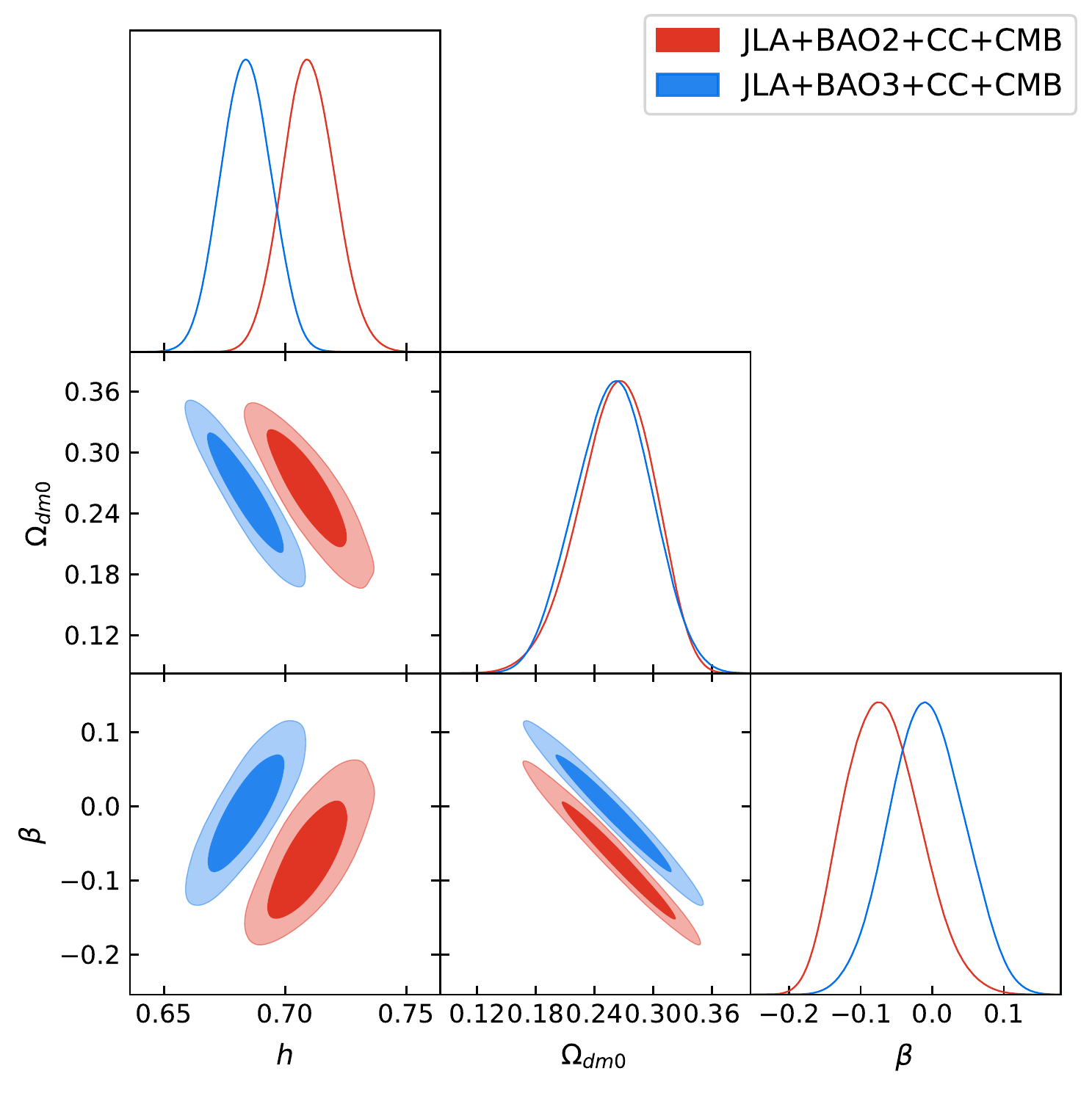}\includegraphics[scale=0.45]{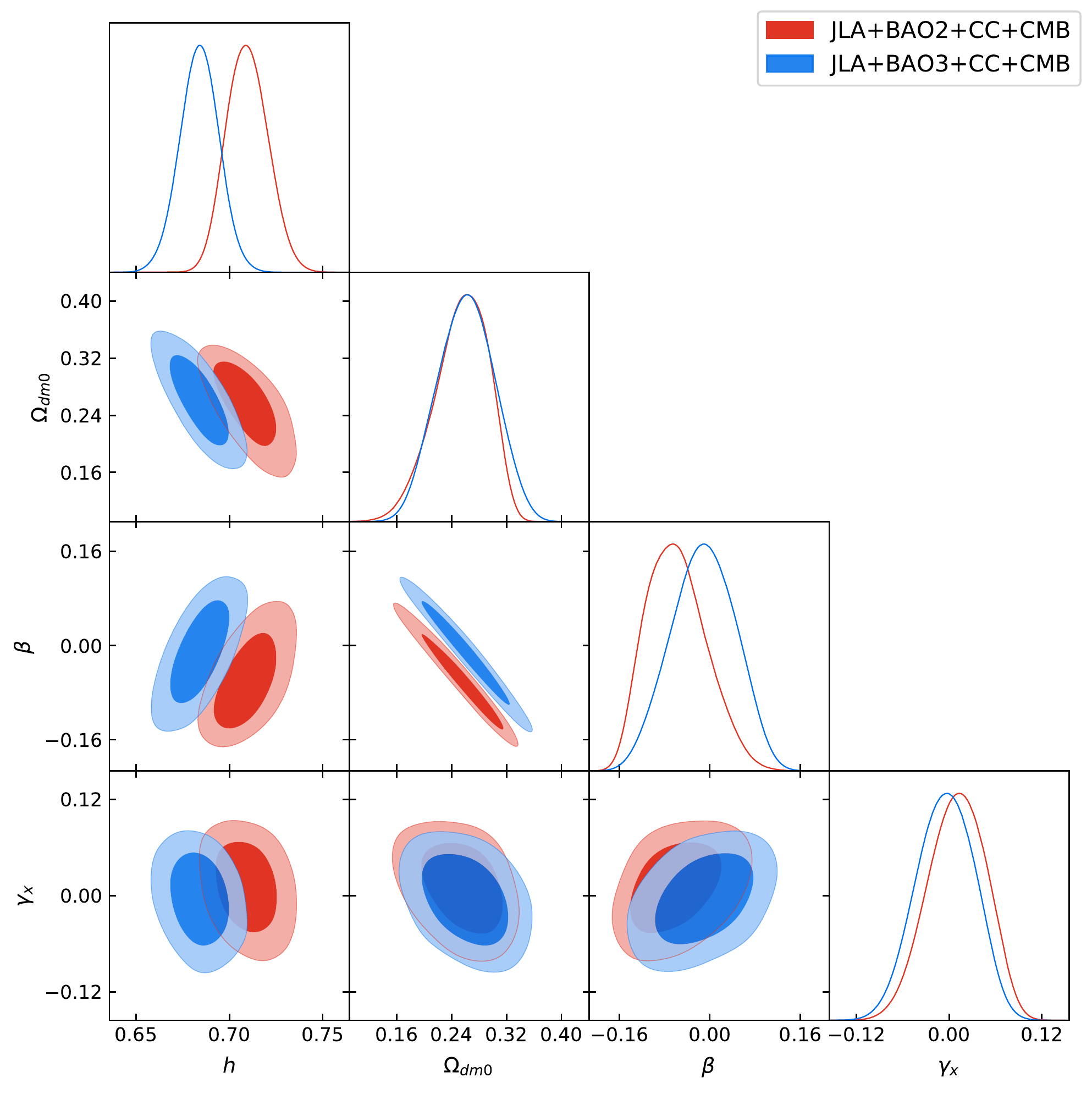}
\caption{\label{contour} Example of contour plots and PDFs (models $\Gamma_{a02}$ and $\Gamma_{a2}$ at left and right, respectively). The results consider as priors scenario \eqref{sc1}.}
\end{figure}

\begin{figure}[ht!]
\centering
\includegraphics[scale=0.32]{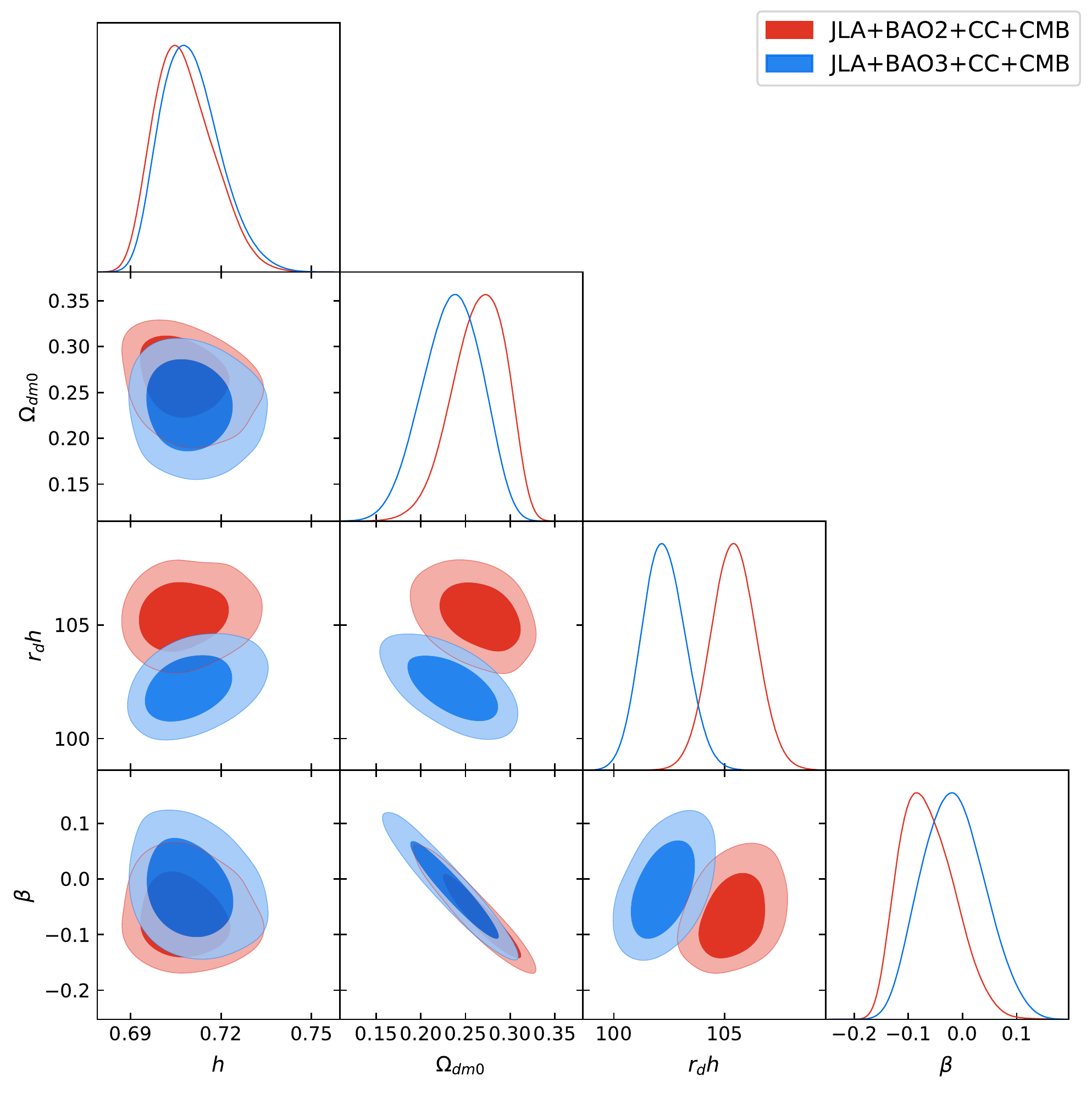}
\includegraphics[scale=0.32]{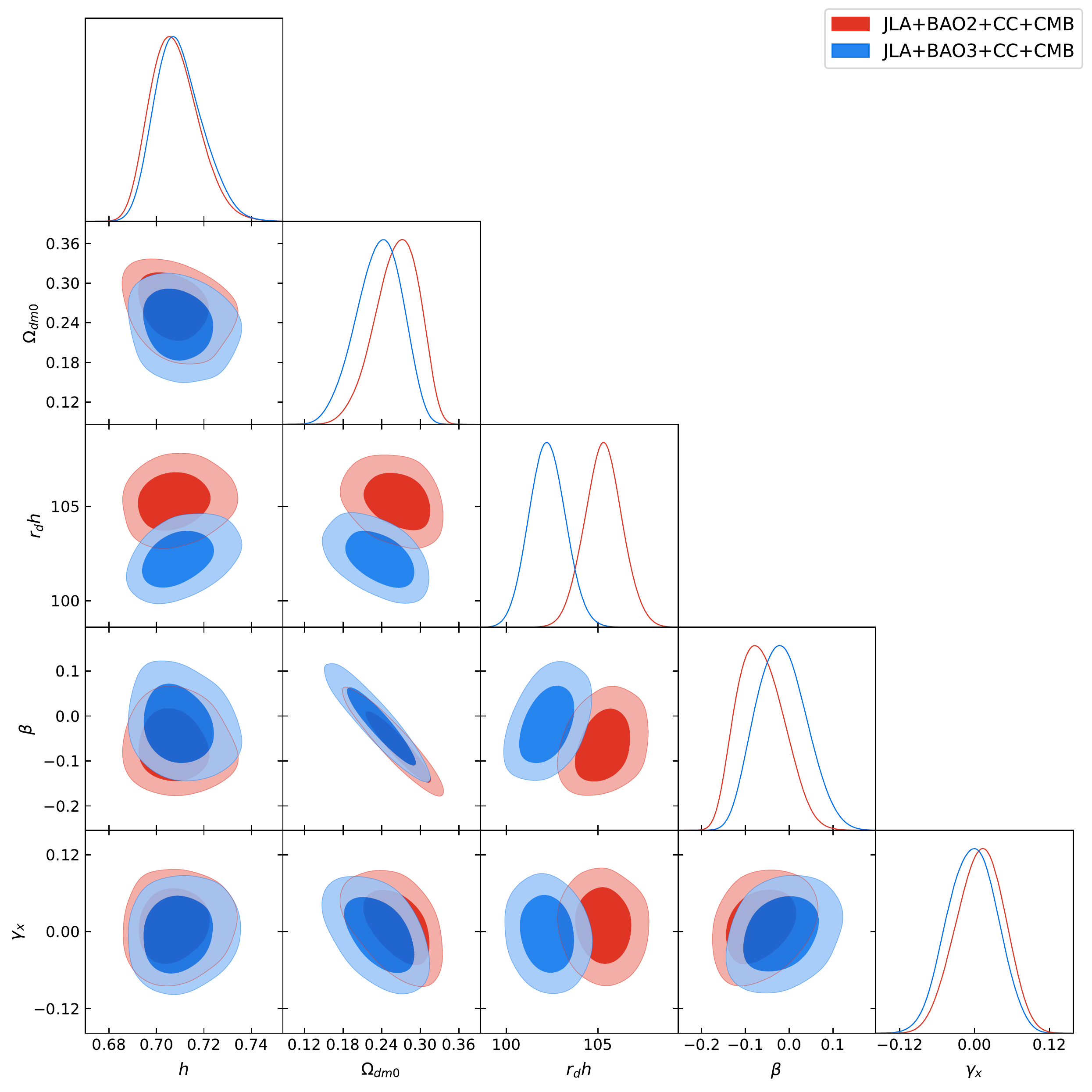}
\caption{\label{c2} Example of contour plots and PDFs (models $\Gamma_{a02}$ and $\Gamma_{a2}$ at left and right, respectively). The results consider as priors scenario \eqref{sc2}.}
\end{figure}

In figure~\ref{comparison} we summarize our results in terms of the interpretation of the Bayes' factor considering the Jeffreys' scale. We show the Bayes' factor for the full joint analysis of the several realizations of interacting models in table \ref{T1} compared to $\Lambda$CDM. We consider independently the full joint analysis using BAO2 and BAO3 data. We see that using BAO3 instead of BAO2 shift the Bayes factor towards a better support for the $\Lambda$CDM model in most of the cases. In figures~\ref{contour} and \ref{c2} we show examples of contour plots and PDFs in our analysis (models $\Gamma_{a02}$ and $\Gamma_{a2}$), for scenarios \eqref{sc1} and \eqref{sc2}  as priors, respectively. The figures show the differences in parameter estimation considering BAO2 or BAO3 in the full joint analysis. In figure \ref{contour} we notice a tension in the parameter estimation using BAO2 or BAO3, this tension is slightly released when we enlarge the parameter space by adding the barotropic index $\gamma_x$ to the analysis. 
In figure \ref{c2} we find that the tension between the parameters estimated is transfered to the $r_dh$ parameter.  The behavior observed in figures \ref{contour} and \ref{c2} is analogous for all the studied scenarios, and therefore other contour plots are not shown for brevity.

In short, our results indicate that the inconclusive evidences obtained for scenarios $\Gamma_{a02}$, $\Gamma_{c0}$, $\Gamma_{e0}$, $\Gamma_{a2}$, $\Gamma_{b2}$, $\Gamma_{c}$ and $\Gamma_{e}$ in comparison to $\Lambda$CDM are maintained in the full joint analysis, this is independent of the chosen priors (scenarios \eqref{sc1} or \eqref{sc2}) or the JLA-Pantheon sample interchange. Nonetheless, in considering $r_dh$ as a free parameter and applying a Gaussian prior to $h$ ($z_*$ in scenario \eqref{sc2} is not relevant in this statement, see Appendix \ref{pantheon} ) it is not so clear that BAO3 data is favoring the $\Lambda$CDM scenario or that BAO2 data is alleviating the $H_0$ tension as it was when considering the scenario \eqref{sc1} as priors. Our results seem to indicate that the BAO3 model-dependency is partially contained in the estimation of $r_d$ as the sound horizon at the drag epoch through \eqref{rs}.

\begin{table}[ht!]
\centering
\caption{\label{Tf1}Bayesian evidence ($\ln \mathcal{E}$) for the models studied in this work. For the analysis with BAO2, CC, JLA and BAO2+CC the evidence is inconclusive for all cases. As for the joint analysis BAO2+CC+CMB, note that $\ln \mathcal{E}_i - \ln \mathcal{E}_{\Lambda \rm{CDM}} <-1$ favors $\Lambda$CDM. These results consider as priors scenario \eqref{sc1}.}
\resizebox{\textwidth}{!}{%
\begin{tabular}{c|c |c |c |c ||c c}
\hline
\multicolumn{1}{c|}{Dataset} & \multicolumn{1}{c|}{BAO2} & \multicolumn{1}{c|}{CC} &\multicolumn{1}{c|}{JLA} &\multicolumn{1}{c||}{BAO2+CC} &\multicolumn{2}{c}{BAO2+CC+CMB} \\\hline\hline
Model & $\ln \mathcal{E}$ & $\ln \mathcal{E}$ & $\ln \mathcal{E}$ & $\ln \mathcal{E}$ &$\ln \mathcal{E}$&Interpretation  \\\hline
$\Lambda$CDM &$-10.221\pm0.008$ & $-7.494\pm0.007$& $-355.493\pm 0.078$& $-17.129\pm0.007$&$-20.617\pm0.018$ & -\\
$\omega$CDM  &$-10.371\pm0.009 $ & $-7.899\pm0.007$&$-355.618 \pm 0.017$& $-17.519\pm0.008$& $-21.061\pm0.012$ & Inconclusive\\
$\Gamma_{a01}$ &$-10.881\pm0.015$& $-7.791\pm0.008$& $-355.645\pm 0.008$&  $-17.557\pm0.007$& $-23.187\pm0.054$ & Moderate\\
$\Gamma_{a02}$ &$-10.268\pm0.010 $& $-7.900\pm0.007$&$-355.641\pm 0.033$ &  $-17.280\pm0.012$& $-20.864\pm0.019$ & Inconclusive \\
$\Gamma_{a03}$ &$-11.085\pm0.009$ & $-7.860\pm0.009$ &$-355.627\pm 0.007$&  $-17.642\pm0.011$& $-23.340\pm0.019$ & Moderate\\
$\Gamma_{a04}$ &$-10.890\pm0.012$ & $-7.907\pm0.008$&$-355.690\pm 0.022$ &$-17.663\pm0.012$& $-23.259\pm0.045$ & Moderate \\
$\Gamma_{b01}$ &$-10.806\pm0.013 $& $-7.896\pm0.008$&$-355.647\pm 0.023$& $-17.468\pm0.015$& $-23.224\pm0.046$ & Moderate\\
$\Gamma_{b03}$ &$-10.948\pm0.009$  & $-7.919\pm0.007$&$-355.566\pm 0.009$ & $-17.541\pm0.009$& $-23.325\pm0.017$ & Weak\\
$\Gamma_{b04}$ &$-10.886\pm0.013$ & $-7.988\pm0.014$&$-355.783\pm 0.018$& $-17.731\pm0.019$ & $-23.208\pm0.074$ & Moderate\\
$\Gamma_{c0}$ &$-10.283\pm0.009$ & $-7.911\pm0.008$ &$-355.634\pm 0.010$&$-17.323\pm0.010$ & $-20.876\pm0.015$ & Inconclusive \\
$\Gamma_{d0}$ &$-10.824\pm0.011$ & $-7.482\pm0.014$& $-355.519\pm 0.033$& $-17.414\pm0.035 $& $-22.954\pm0.106$ & Weak\\
$\Gamma_{e0}$ &$-10.290\pm0.009$ & $-7.895\pm0.008$& $ -355.596\pm 0.021$& $-17.407\pm0.007$ & $-20.812\pm0.055$ & Inconclusive\\
$\Gamma_{a1}$ &$-10.905\pm0.014$ & $-7.959\pm0.008$&$-355.692\pm 0.039$ &$-17.598\pm0.186$& $-22.338\pm0.481$ & Weak\\
$\Gamma_{a2}$ &$-10.355\pm0.011$ & $-7.918\pm0.013$& $-355.707\pm 0.021$&$-17.306\pm0.014$& $-20.889\pm0.044$ & Inconclusive \\
$\Gamma_{a3}$ &$-10.988\pm0.014$ & $-7.851\pm0.009$& $-355.694\pm 0.087$&$-17.784\pm0.036$& $-23.232\pm0.061$ & Moderate\\
$\Gamma_{a4}$ &$-10.900\pm0.015$ & $-7.866\pm0.011$& $-355.543\pm 0.208$&$-17.712\pm0.015$& $-23.240\pm0.082$ & Moderate \\
$\Gamma_{b1}$ &$-10.879\pm0.012$ & $-7.909\pm0.008$& $-355.727\pm 0.026$&$-17.798\pm0.014$ & $-23.160\pm0.093$ & Moderate \\
$\Gamma_{b2}$ &$-10.439\pm0.012$ & $-8.039\pm0.011$& $ -355.468\pm 0.227$&$-17.544\pm0.008$& $-20.907\pm0.080$ & Inconclusive\\
$\Gamma_{b3}$ &$-10.962\pm0.012$ & $-7.979\pm0.009$& $-355.755\pm 0.008$&$-17.859\pm0.010$ & $-23.053\pm0.108$ & Weak\\
$\Gamma_{b4}$ &$-10.898\pm0.013$ & $-7.979\pm0.009$& $-355.529\pm 0.318$&$-17.534\pm0.047$ & $-23.059\pm0.093$ &Weak \\
$\Gamma_{c}$ &$-10.387\pm0.009$ & $-8.033\pm0.008$& $-355.688\pm 0.014$&$-17.332\pm0.023$ & $-20.793\pm0.054$  & Inconclusive \\
$\Gamma_{d}$ &$-10.825\pm0.015$ & $-8.004\pm0.010$& $-355.647\pm 0.019$&$-17.600\pm0.042$& $-22.749\pm0.369$  & Weak \\
$\Gamma_{e}$ &$-10.386\pm0.009$ &$-7.998\pm0.100$& $-355.799\pm 0.009$&$-17.369\pm0.009$ & $-20.881\pm0.051$  & Inconclusive \\\hline
\end{tabular}}
\end{table}

\begin{table}[ht!]
\centering
\caption{\label{Tf2} Bayesian evidence ($\ln \mathcal{E}$) and interpretation for the models considered in this work. Note that $\ln B_i = \ln \mathcal{E}_i - \ln \mathcal{E}_{\Lambda \rm{CDM}} <-1$ favors the $\Lambda$CDM model. These results consider as priors scenario \eqref{sc1}.}
\resizebox{\textwidth}{!}{%
\begin{tabular}{c|c c|c c|c c}\hline
\multicolumn{1}{c|}{Dataset} & \multicolumn{2}{c|}{BAO3} & \multicolumn{2}{c|}{BAO3 + CC} &\multicolumn{2}{c}{BAO3 + CC + CMB} \\\hline\hline
Model & $\ln \mathcal{E}$  & Interpretation & $\ln \mathcal{E}$   & Interpretation & $\ln \mathcal{E}$  & Interpretation \\\hline
$\Lambda$CDM   &$-5.218\pm0.007$  & -& $-11.167\pm0.008$& - &$-14.673\pm0.008$ &-\\
$\omega$CDM    &$-5.559\pm0.008$ & Inconclusive &
$-11.630\pm0.008$& Inconclusive&$-14.631\pm0.206$ &Inconclusive\\
$\Gamma_{a01}$ &$-7.037\pm0.015$ & Weak &
$-12.894\pm0.060$& Weak& $-17.792\pm0.031$ &Moderate\\
$\Gamma_{a02}$ &$-5.417\pm0.008$ & Inconclusive&
$-11.437\pm0.008$&Inconclusive&$-14.770\pm0.013 $&Inconclusive\\
$\Gamma_{a03}$ &$-7.036\pm0.030$  & Weak&
$-13.008\pm0.027$ & Weak&$-17.824\pm0.016$&Moderate\\
$\Gamma_{a04}$ &$-6.918\pm0.018$ &  Weak&
$-13.083\pm0.014$ & Weak&$-17.663\pm0.100$ &Moderate\\
$\Gamma_{b01}$ &$-6.910\pm0.055$ & Weak&
$-12.974\pm0.009$ & Weak&$-17.835\pm0.018 $&Moderate\\
$\Gamma_{b03}$ &$-6.976\pm0.019$ & Weak&
$-12.977\pm0.008$ &Weak&$-17.788\pm0.027$ &Moderate\\
$\Gamma_{b04}$ &$-6.999\pm0.036$ &  Weak&
$-13.354\pm0.037$& Weak&$-17.081\pm0.373$ &Weak\\
$\Gamma_{c0}$  &$-5.458\pm0.010$ & Inconclusive&
$-11.439\pm0.010$& Inconclusive&$-14.748\pm0.015$&Inconclusive\\
$\Gamma_{d0}$  &$-7.004\pm0.016$ & Weak&
$-12.929\pm0.008$ & Weak&$-17.803\pm0.096$ &Moderate\\
$\Gamma_{e0}$  &$-5.441\pm0.008$  & Inconclusive&
$-11.472\pm0.008$ & Inconclusive&$-14.360\pm0.264$ &Inconclusive\\
$\Gamma_{a1}$  &$-6.979\pm0.026$ &  Weak&
$-13.203\pm0.016$ & Weak&$-17.801\pm0.024$ &Moderate\\
$\Gamma_{a2}$  &$-5.466\pm0.012$ &  Inconclusive&
$-11.473\pm0.015$ & Inconclusive&$-14.759\pm0.021$ &Inconclusive\\
$\Gamma_{a3}$  &$-7.105\pm0.012$ &  Weak&
$-13.165\pm0.101$ & Weak&$-17.644\pm0.111$ &Moderate\\
$\Gamma_{a4}$  &$-6.974\pm0.026$ & Weak&
$-13.143\pm0.012$&  Weak&$-17.477\pm0.286$ &Moderate\\
$\Gamma_{b1}$  &$-6.974\pm0.013$ &Weak&
$-13.194\pm0.035$ & Weak&$-17.734\pm0.044$ &Moderate\\
$\Gamma_{b2}$  &$-5.596\pm0.012$ & Inconclusive&
$-11.794\pm0.008$ &Inconclusive&$-14.621\pm0.115 $&Inconclusive\\
$\Gamma_{b3}$  &$-7.049\pm0.013$ & Weak&
$-13.219\pm0.017$ & Weak&$-17.605\pm0.145$ &Moderate\\
$\Gamma_{b4}$  &$-6.985\pm0.021$ & Weak&
$-13.269\pm0.013$ & Weak&$-17.322\pm0.260$ &Moderate\\
$\Gamma_{c}$   &$-5.515\pm0.008$ &  Inconclusive&
$-11.479\pm0.011$ & Inconclusive&$-14.731\pm0.013$ &Inconclusive\\
$\Gamma_{d}$   &$-7.026\pm0.012$ & Weak&
$-13.13\pm0.037$ & Weak&$-17.760\pm0.037$&Moderate\\
$\Gamma_{e}$   &$-5.534\pm0.009$  & Inconclusive&
$-11.568\pm0.008$ & Inconclusive&$-14.691\pm0.052$&Inconclusive\\
\hline
\end{tabular}}
\end{table}

\begin{landscape}
\begin{table}[ht!]
\centering
\caption{\label{full}Best fit parameters for the joint analysis JLA + BAO2 + CC + CMB. The last three columns show the Bayesian evidence ($\ln \mathcal{E}$), the Bayes factor ($\ln B$) and the interpretation of the strength of the evidence. Note that $\ln B<-1$ favors the $\Lambda$CDM model. These results consider as priors scenario \eqref{sc1}. }
\resizebox{1.5\textwidth}{!}{%
{\renewcommand{\arraystretch}{1.4}%
\begin{tabular}{c|cccccccc}\hline
Model & $h$ & $\Omega_{dm0}$  & $\alpha$ & $\beta$&$\gamma_x$&$\ln \mathcal{E}$&$\ln B$&Interpretation\\\hline
$\Lambda$CDM & $0.720\pm0.008$ & $0.2127^{+0.0078}_{-0.0088}$ &-&-&-&$-374.379\pm0.006$& 0 &-\\
$\omega$CDM & $0.717\pm 0.009$ & $0.2119^{+0.0078}_{-0.0087}$ &-&-& $0.023^{+0.037}_{-0.028}$ & $-374.120\pm0.072$ & $0.259\pm0.072$&Inconclusive\\
$\Gamma_{a01}$ & $0.693\pm 0.020$& $0.2166^{+0.0084}_{-0.0094}$& $-0.0077^{+0.0065}_{-0.0045}$ &-&-&$-376.557\pm0.063$&  $-2.178\pm0.063$&Weak\\
$\Gamma_{a02}$ & $0.710\pm 0.011$ & $0.262^{+0.041}_{-0.034}   $ &-& $-0.069^{+0.047}_{-0.057}  $ &-& $-373.830\pm0.160$& $0.549\pm0.160$&Inconclusive\\
$\Gamma_{a03}$ & $0.691\pm0.021$ & $0.222^{+0.010}_{-0.012}   $ & $-0.0079^{+0.0064}_{-0.0047}$ & $-0.0079^{+0.0064}_{-0.0047}$ &-&$-375.824\pm0.290$ & $-1.445\pm0.290$&Weak\\
$\Gamma_{a04}$ & $0.687\pm0.020$ & $0.259^{+0.042}_{-0.034}$ & $-0.0065^{+0.0061}_{-0.0041}$ & $-0.060^{+0.048}_{-0.058}$ &-& $-376.222\pm0.135$ &$-1.843\pm0.135$&Weak\\
$\Gamma_{b01}$ & $0.692\pm 0.021$  & $0.2167^{+0.0084}_{-0.0096}$ &$0.0079^{+0.0047}_{-0.0068}$  &-&-&$-376.332\pm0.259$ & $-1.953\pm0.259$&Weak\\
$\Gamma_{b03}$ & $0.693\pm 0.021$ & $0.2165^{+0.0085}_{-0.0098}$ & $0.0078^{+0.0046}_{-0.0065}$ &$0.0078^{+0.0046}_{-0.0065}$ &-& $-376.468\pm0.081$&$-2.089\pm0.081$&Weak\\
$\Gamma_{b04}$ & $0.693\pm0.021$ &$0.2165^{+0.0086}_{-0.0096}$ & $0.0078^{+0.0046}_{-0.0071}$ & $-0.012\pm0.097$&-& $-376.587\pm0.107$&$-2.208\pm0.107$&Weak\\
$\Gamma_{c0}$ & $0.7134\pm 0.0099$& $0.237\pm 0.024 $& $-0.075^{+0.057}_{-0.079}$&-&-& $-374.004\pm0.028$&$0.375\pm0.029$&Inconclusive\\
$\Gamma_{d0}$ & $0.694\pm 0.021$&$0.2142^{+0.0081}_{-0.0091}$ & $-0.0076^{+0.0068}_{-0.0047}$&-&-& $-376.276\pm0.224$&$-1.897\pm0.224$&Weak\\
$\Gamma_{e0}$ &$0.7173\pm 0.0086$ & $0.232^{+0.033}_{-0.024}$&$-0.048^{+0.059}_{-0.075}$ &-&-&$-374.049\pm0.184$ &$0.330\pm0.184$&Inconclusive\\
$\Gamma_{a1}$ &$0.690\pm 0.021$ & $0.2159^{+0.0084}_{-0.0094}$	&$-0.0076^{+0.0066}_{-0.0046}$ &-& $0.023^{+0.039}_{-0.032}   $& $-376.373\pm0.171$&$-1.994\pm0.171$&Weak\\
$\Gamma_{a2}$ & $0.710\pm 0.011$& $0.255^{+0.045}_{-0.032}$&-&$-0.059^{+0.044}_{-0.062}	$ &$0.011^{+0.040}_{-0.036}$ &$-374.116\pm0.057$ &$0.263\pm0.057$&Inconclusive\\
$\Gamma_{a3}$ & $0.689\pm 0.021$&$0.2214^{+0.0098}_{-0.012}$ &$-0.0075^{+0.0063}_{-0.0046}$ & $-0.0075^{+0.0063}_{-0.0046}$ &$0.020^{+0.038}_{-0.031}   $ & $-376.375\pm0.150$&$-1.996\pm0.150$&Weak\\
$\Gamma_{a4}$ & $0.688\pm 0.020$ & $0.252^{+0.045}_{-0.038}$ & $-0.0064^{+0.0060}_{-0.0042}$ & $-0.051^{+0.053}_{-0.061}$ & $0.011\pm 0.039$ & $-376.557\pm0.037$ &$-2.178\pm0.037$&Weak\\
$\Gamma_{b1}$ &$0.691\pm 0.021 $ & $0.2159\pm 0.0090$&$0.0074^{+0.0046}_{-0.0066}$ &-& $0.024^{+0.039}_{-0.033}$& $-376.482\pm0.047$&$-2.103\pm0.047$&Weak\\
$\Gamma_{b2}$ & $0.7164\pm 0.0089$ & $0.2124\pm 0.0085$&-& $0.000\pm 0.095$ & $0.025^{+0.037}_{-0.029}$ &$-374.418\pm0.020$ &$-0.039\pm0.021$&Inconclusive\\
$\Gamma_{b3}$ &$0.691\pm 0.021$ &$0.2160^{+0.0084}_{-0.0097}$ & $0.0074^{+0.0047}_{-0.0064}$&$0.0074^{+0.0047}_{-0.0064}$ & $0.023^{+0.039}_{-0.031}$& $-376.492\pm0.050$&$-2.113\pm0.050$&Weak\\
$\Gamma_{b4}$ & $0.690\pm 0.020$ &$0.2159\pm 0.0093$ & $0.0075^{+0.0045}_{-0.0065}$ & $-0.007\pm 0.094$ &$0.024^{+0.038}_{-0.032}$ & $-376.670\pm0.040$&$-2.291\pm0.040$&Weak\\
$\Gamma_{c}$ &$0.711\pm 0.010$ &$0.233^{+0.026}_{-0.023}$ & $-0.065^{+0.057}_{-0.082}$ &-& $0.019^{+0.038}_{-0.032}$ & $-374.118\pm0.022$&$0.261\pm0.023$&Inconclusive\\
$\Gamma_{d}$ & $0.692\pm 0.020$ & $0.2134^{+0.0080}_{-0.0089}$ & $-0.0073^{+0.0065}_{-0.0046}$&-& $0.024^{+0.038}_{-0.031}$ & $-376.575\pm0.044$ &$-2.196\pm0.044$&Weak\\
$\Gamma_{e}$ &$0.7143\pm 0.0094$ & $0.227^{+0.034}_{-0.023}$ & $-0.036^{+0.056}_{-0.076} $ 	&-& $0.024^{+0.039}_{-0.031}$ & $-374.271\pm0.103$ &$0.108\pm0.103$&Inconclusive\\
\hline
\end{tabular}}}
\end{table}
\end{landscape}

\begin{landscape}
\begin{table}[ht!]
\centering
\caption{\label{full3}Best fit parameters for the joint analysis JLA + BAO3 + CC + CMB. The last three columns show the Bayesian evidence ($\ln \mathcal{E}$), the Bayes factor ($\ln B$) and the interpretation of the strength of the evidence. Note that $\ln B<-1$ favors the $\Lambda$CDM model. These results consider as priors scenario  \eqref{sc1}.}
\resizebox{1.5\textwidth}{!}{%
{\renewcommand{\arraystretch}{1.4}%
\begin{tabular}{c|cccccccc}
\hline
Model & $h$ & $\Omega_{dm0}$  & $\alpha$ & $\beta$&$\gamma_x$&$\ln \mathcal{E}$&$\ln B$&Interpretation\\\hline
$\Lambda$CDM & $0.6853\pm 0.0061$ & $0.2537\pm 0.0080$ &-&-&-&$-366.953\pm0.005$& - &\\
$\omega$CDM & $0.6854\pm 0.0081$ & $0.2536\pm 0.0080$ &-&-& $-0.001\pm 0.035$ & $-367.254\pm0.017$ & $-0.301\pm0.018$ &Inconclusive\\
$\Gamma_{a01}$ & $0.684\pm 0.020$& $0.2543\pm 0.0088$& $-0.0005^{+0.0044}_{-0.0035}$ &-&-&$-370.216\pm0.114$& $-3.263\pm0.114$ &Moderate\\
$\Gamma_{a02}$ & $0.684\pm 0.010$ & $0.259\pm 0.038$ &-& $-0.008\pm 0.052$ &-& $-367.441\pm0.114$& $-0.488\pm0.114$&Inconclusive\\
$\Gamma_{a03}$ & $0.684\pm 0.020$ & $0.255\pm 0.010$ & $-0.0005^{+0.0044}_{-0.0035}$ & $-0.0005^{+0.0044}_{-0.0035}$ &-&$-370.003\pm0.150$ & $-3.050\pm0.150$&Moderate\\
$\Gamma_{a04}$ & $0.683\pm 0.020$ & $0.256\pm 0.039$& $-0.0006^{+0.0045}_{-0.0034}$ & $-0.003\pm 0.053$ &-& $-370.615\pm0.194$ &$-3.662\pm0.194$&Moderate\\
$\Gamma_{b01}$ & $0.684\pm 0.020$  & $0.2543\pm 0.0090$ &$0.0006^{+0.0035}_{-0.0045}$ &-&-&$-370.091\pm0.157$ & $-3.138\pm0.157$&Moderate\\
$\Gamma_{b03}$ & $0.684\pm 0.019$ & $0.2542^{+0.0083}_{-0.0094}$ & $0.0005^{+0.0034}_{-0.0043}$ &$0.0005^{+0.0034}_{-0.0043}$ &-& $-370.286\pm0.048$&$-3.333\pm0.048$&Moderate\\
$\Gamma_{b04}$ & $0.684\pm 0.019$ &$0.2542\pm 0.0087$ & $0.0005^{+0.0034}_{-0.0043}$ & $-0.011\pm 0.098$ &-& $-370.193\pm0.120$&$-3.240\pm0.120$&Moderate\\
$\Gamma_{c0}$ & $0.6839\pm 0.0091$& $0.258\pm 0.025$& $-0.011\pm 0.068$ &-&-& $-367.110\pm0.194$&$-0.157\pm0.194$&Inconclusive\\
$\Gamma_{d0}$ & $0.684\pm 0.020$&$0.2542\pm 0.0086$& $-0.0007^{+0.0046}_{-0.0035}$&-&-& $-370.213\pm0.058$&$-3.260\pm0.058$&Moderate\\
$\Gamma_{e0}$ &$0.6852\pm 0.0075$ & $0.254^{+0.032}_{-0.027}$ &$-0.002\pm 0.072$ &-&-&$-367.044\pm0.078$ &$-0.091\pm0.078$&Inconclusive\\
$\Gamma_{a1}$ &$0.684\pm 0.020$ & $0.2544\pm 0.0088$ &$-0.0006^{+0.0044}_{-0.0035}$ &-& $0.000\pm 0.035$ & $-370.457\pm0.131$&$-3.504\pm0.131$&Moderate\\
$\Gamma_{a2}$ & $0.684\pm 0.010$& $0.261\pm 0.041$ &-& $-0.011\pm 0.056 $ &$-0.004\pm 0.037$ &$-367.683\pm0.087$ &$-0.730\pm0.087$&Inconclusive\\
$\Gamma_{a3}$ & $0.684\pm 0.020$&$0.2547^{+0.0095}_{-0.011} $ &$-0.0005^{+0.0044}_{-0.0034}$ & $-0.0005^{+0.0044}_{-0.0034}$ & $0.001\pm 0.036$ & $-369.561\pm0.642$&$-2.608\pm0.642$&Moderate\\
$\Gamma_{a4}$ & $0.683\pm 0.020$ & $0.256^{+0.044}_{-0.039} $ & $-0.0006^{+0.0045}_{-0.0033}$ & $-0.004\pm 0.057 $ & $-0.002\pm 0.038$ & $-370.905\pm0.039$ &$-3.952\pm0.039$&Moderate\\
$\Gamma_{b1}$ &$0.683\pm 0.020 $ & $0.2545\pm 0.0088$ & $0.0006^{+0.0034}_{-0.0045}$ &-& $0.001\pm 0.035 $& $-370.509\pm0.055$&$-3.556\pm0.055$&Moderate\\
$\Gamma_{b2}$ & $0.6853\pm 0.0080$ &  $0.2539\pm 0.0082$ &-& $-0.005\pm 0.093$ & $-0.001\pm 0.034$ &$-367.196\pm0.030$ &$-0.243\pm0.030$&Inconclusive\\
$\Gamma_{b3}$ &$0.684\pm 0.020$ &$0.2542\pm 0.0089$ & $0.0005^{+0.0034}_{-0.0044}$ &$0.0005^{+0.0034}_{-0.0044}$ & $0.000\pm 0.036$& $-370.416\pm0.093$ &$-3.463\pm0.093$&Moderate\\
$\Gamma_{b4}$ & $0.684\pm 0.020 $ & $0.2543\pm 0.0092$ & $0.0007^{+0.0035}_{-0.0045}$ & $-0.016\pm 0.094$ &$0.000\pm 0.036$ & $-370.560\pm0.030$&$-3.607\pm0.030$&Moderate\\
$\Gamma_{c}$ & $0.6844\pm 0.0097$ &  $0.258\pm 0.026$ & $-0.011\pm 0.072$ &-& $-0.002\pm 0.036$ & $-367.189\pm0.208$&$-0.236\pm0.208$&Inconclusive\\
$\Gamma_{d}$ & $0.684\pm 0.020$ & $0.2543\pm 0.0084$ & $-0.0006^{+0.0045}_{-0.0035}$&-& $0.000\pm 0.036$ & $-370.328\pm0.184$ &$-3.375\pm0.184$&Moderate\\
$\Gamma_{e}$ &$0.6851\pm 0.0086$ & $0.254^{+0.031}_{-0.026}$ & $-0.004\pm 0.069$ 	&-& $0.000\pm0.035$ & $-367.306\pm0.070$ &$-0.353\pm0.070$&Inconclusive\\
\hline
\end{tabular}}}
\end{table}
\end{landscape}

\begin{landscape}
\begin{table}[ht!]
\centering
\caption{\label{full_nJB2}Best fit parameters for the joint analysis JLA + BAO2 + CC + CMB. The last three columns show the Bayesian evidence ($\ln \mathcal{E}$), the Bayes factor ($\ln B$) and the interpretation of the strength of the evidence. Note that $\ln B<-1$ favors the $\Lambda$CDM model. These results consider as priors scenario \eqref{sc2}. }
\resizebox{1.5\textwidth}{!}{%
{\renewcommand{\arraystretch}{1.5}%
\begin{tabular}{c|ccccccccc}\hline
Model & $h$ & $\Omega_{dm0}$  &$r_dh$ [Mpc]& $\alpha$ & $\beta$&$\gamma_x$&$\ln \mathcal{E}$&$\ln B$&Interpretation\\\hline
$\Lambda$CDM & $0.7054^{+0.0072}_{-0.012}$ & $0.231^{+0.014}_{-0.0092}$ &$105.73\pm 0.98$ &-&-&-&$-375.502\pm0.045$& -&\\
$\omega$CDM & $0.7058^{+0.0076}_{-0.011}$  & $0.226^{+0.015}_{-0.012}$  &$105.6\pm 1.0$   &-&-& $0.020^{+0.039}_{-0.032}$ & $-374.785\pm0.418$ & $0.717\pm0.420$&Inconclusive\\
$\Gamma_{a01}$ & $0.7104^{+0.0094}_{-0.013}$& $0.258\pm 0.024$&$105.2\pm 1.1$& $0.0051^{+0.0038}_{-0.0026}$ &-&-&$-377.724\pm0.074$& $-2.222\pm0.087$ &Weak\\
$\Gamma_{a02}$ & $0.7079^{+0.0079}_{-0.012}$ & $0.265^{+0.034}_{-0.025}$ &$105.4\pm 1.0$&-& $-0.064^{+0.041}_{-0.059}$ &-& $-375.428\pm0.054$&$0.074\pm0.070$&Inconclusive\\
$\Gamma_{a03}$ & $0.7091^{+0.0096}_{-0.012} $ &$0.286\pm 0.042$ &$105.1\pm 1.1$& $-0.052^{+0.063}_{-0.091}$ & $-0.052^{+0.063}_{-0.091}$ &-&$-376.625\pm0.314$ &$-1.123\pm0.317$ &Weak\\
$\Gamma_{a04}$ &$0.709^{+0.011}_{-0.013}$ &  $0.266\pm 0.034$ &$105.1\pm 1.1$& $0.0030^{+0.0061}_{-0.0016}$ & $-0.015^{+0.066}_{-0.084}  $ &-& $-377.429\pm0.179$& $-1.927\pm0.184$ &Weak \\
$\Gamma_{b01}$ & $0.709^{+0.010}_{-0.013}$  & $0.259\pm 0.023$ &$105.2\pm 1.1$&$-0.0050^{+0.0025}_{-0.0037}$  &-&-&$-377.719\pm0.047$ & $-2.217\pm0.065$&Weak\\
$\Gamma_{b03}$ & $0.7103^{+0.0092}_{-0.013} $ & $0.258\pm 0.023$ &$105.2\pm 1.1$& $-0.0051^{+0.0026}_{-0.0038}$ &$-0.0051^{+0.0026}_{-0.0038}$ &-& $-377.710\pm0.048$&$-2.208\pm0.066$&Weak\\
$\Gamma_{b04}$ & $0.7102^{+0.0090}_{-0.013} $ & $0.260\pm 0.022$ &$105.2\pm 1.1$& $-0.0053^{+0.0027}_{-0.0035}$ & $-0.012\pm 0.095$&-& $-377.796\pm0.064$&$-2.294\pm0.078$&Weak\\
$\Gamma_{c0}$ & $0.7065^{+0.0074}_{-0.011}$& $0.242\pm 0.019$&$105.59\pm 0.99$&$-0.054^{+0.062}_{-0.083}$&-&-& $-375.515\pm0.081$&$-0.013\pm0.093$&Inconclusive\\
$\Gamma_{d0}$ & $0.7104^{+0.0095}_{-0.013} $&$0.259\pm 0.024$ &$105.2\pm 1.1$& $0.0051^{+0.0035}_{-0.0026}$&-&-& $-377.830\pm0.040$&$-2.328\pm0.060$&Weak\\
$\Gamma_{e0}$ &$0.7066^{+0.0078}_{-0.011} $& $0.248^{+0.028}_{-0.021}$&$105.57\pm 0.98$ &$-0.055^{+0.055}_{-0.078}$ &-&-&$-375.368\pm0.192$ &$0.134\pm0.197$&Inconclusive\\
$\Gamma_{a1}$ &$0.7097^{+0.0099}_{-0.012} $& $0.258\pm 0.027$&$105.2\pm 1.1$ &$0.0050^{+0.0038}_{-0.0026}$ &-& $0.004\pm 0.040$& $-377.870\pm0.050$&$-2.368\pm0.067$&Weak\\
$\Gamma_{a2}$ & $0.7077^{+0.0085}_{-0.011} $& $0.264^{+0.039}_{-0.029}$&$105.3\pm 1.0$&-&$-0.065^{+0.046}_{-0.060}$ &$0.0096\pm 0.038$ &$-375.240\pm0.215$ &$0.262\pm0.220$&Inconclusive\\
$\Gamma_{a3}$ & $0.709^{+0.010}_{-0.012}$&$0.273^{+0.028}_{-0.045}$ &$105.1\pm 1.0$& $-0.034^{+0.045}_{-0.032}$ & $-0.034^{+0.045}_{-0.032}$ &$0.005\pm 0.040$ & $-377.449\pm0.032	$&$-1.947\pm0.055$&Weak\\
$\Gamma_{a4}$ &$0.709^{+0.010}_{-0.012}$ & $0.264\pm 0.037$ &$105.1\pm 1.0$& $0.0041^{+0.0050}_{-0.0029}$ & $-0.014^{+0.068}_{-0.082}$  &$0.003\pm 0.041$ & $-377.878\pm0.058$ &$-2.376\pm0.073$&Weak\\
$\Gamma_{b1}$ &$0.7099^{+0.0098}_{-0.013} $ & $0.257\pm 0.028$&$105.2\pm 1.0$&$-0.0049^{+0.0025}_{-0.0041}$ &-& $0.004\pm 0.042$& $-377.981\pm0.020$&$-2.479\pm0.049$&Weak\\
$\Gamma_{b2}$ & $0.7064^{+0.0074}_{-0.011}$ & $0.226^{+0.014}_{-0.011}$&$105.62\pm 0.98$&-& $-0.001\pm 0.092$ & $0.020^{+0.038}_{-0.030}$ &$-375.696\pm0.138$ &$-0.194\pm0.145$&Weak\\
$\Gamma_{b3}$ &$0.7092^{+0.0099}_{-0.013} $ &$0.258\pm 0.028 $ &$105.2\pm 1.0$& $-0.0046^{+0.0023}_{-0.0043}$&$-0.0046^{+0.0023}_{-0.0043}$ & $0.003\pm 0.042$& $-376.849\pm0.583$&$-1.347\pm0.585$&Weak\\
$\Gamma_{b4}$ & $0.7091^{+0.0099}_{-0.013} $ &$0.259\pm 0.026$ &$105.2\pm 1.1$&  $-0.0051^{+0.0026}_{-0.0036}$ & $-0.012\pm 0.095$ &$0.003\pm 0.041$& $-377.936\pm0.119$&$-2.434\pm0.127$&Weak\\
$\Gamma_{c}$ & $0.7061^{+0.0076}_{-0.011} $ &$0.238^{+0.023}_{-0.021}$ &$105.47\pm 0.99$& $-0.051^{+0.062}_{-0.085}$ &-& $0.017\pm 0.037$ & $-375.585\pm0.104$&$-0.083\pm0.113$&Inconclusive\\
$\Gamma_{d}$ & $0.7089^{+0.0098}_{-0.013}$ &  $0.259\pm 0.028$ &$105.2\pm 1.0$& $0.0048^{+0.0040}_{-0.0025}$&-& $0.004\pm 0.040$ & $-377.793\pm0.123$ &$-2.291\pm0.131$&Weak\\
$\Gamma_{e}$ &$0.7069^{+0.0080}_{-0.011} $ &  $0.240^{+0.030}_{-0.021}$ &$105.5\pm 1.0$& $-0.044^{+0.053}_{-0.074}$ 	&-& $0.018^{+0.040}_{-0.035}$ & $-375.744\pm0.061$ &$-0.242\pm0.076$&Inconclusive\\
\hline
\end{tabular}}}
\end{table}
\end{landscape}

\begin{landscape}
\begin{table}[ht!]
\centering
\caption{\label{full_nJB3}Best fit parameters for the joint analysis JLA + BAO3 + CC + CMB. The last three columns show the Bayesian evidence ($\ln \mathcal{E}$), the Bayes factor ($\ln B$) and the interpretation of the strength of the evidence. Note that $\ln B<-1$ favors the $\Lambda$CDM model. These results consider as priors scenario \eqref{sc2}.}
\resizebox{1.5\textwidth}{!}{%
{\renewcommand{\arraystretch}{1.5}%
\begin{tabular}{c|ccccccccc}\hline
Model & $h$ & $\Omega_{dm0}$  &$r_dh$ [Mpc]& $\alpha$ & $\beta$&$\gamma_x$&$\ln \mathcal{E}$&$\ln B$&Interpretation\\\hline
$\Lambda$CDM & $0.7100^{+0.0077}_{-0.011}$ & $0.225^{+0.012}_{-0.0095}$ &$102.45\pm 0.88$ &-&-&-&$-366.978\pm0.112$&-&\\
$\omega$CDM &$0.7093^{+0.0080}_{-0.011} $ & $0.226\pm 0.014$& $102.40\pm 0.90$&-&-& $0.000\pm 0.037$ &$-367.306\pm0.020$ &$-0.328\pm0.114$ &Inconclusive\\
$\Gamma_{a01}$ & $0.713^{+0.010}_{-0.012}$& $0.239\pm 0.019$&$101.9\pm 1.0 $& $0.00090^{+0.0060}_{-0.00092}$ &-&-&$-369.887\pm0.097$&  $-2.909\pm0.148$&Moderate\\
$\Gamma_{a02}$ & $0.7100^{+0.0079}_{-0.011} $ & $0.235^{+0.036}_{-0.031}$ &$102.26\pm 0.96$&-& $-0.017^{+0.054}_{-0.062}$&-& $-367.426\pm0.084$&$-0.448\pm0.140$&Inconclusive\\
$\Gamma_{a03}$ & $0.713^{+0.010}_{-0.013}$ &$0.242^{+0.015}_{-0.025}$ &$101.9\pm 1.1$& $-0.0065^{+0.014}_{-0.0059}$ & $-0.0065^{+0.014}_{-0.0059}$ &-&$-369.780\pm0.086$ &$-2.802\pm0.141$ &Moderate\\
$\Gamma_{a04}$ &$0.713^{+0.011}_{-0.013}   $&  $0.232\pm 0.034$ &$101.9\pm 1.1$& $0.0030^{+0.0051}_{-0.0029}$ & $0.020\pm 0.073$ &-& $-369.915\pm0.135$&$-2.937\pm0.175$ &Moderate\\
$\Gamma_{b01}$ & $0.7136^{+0.0099}_{-0.013}$ & $0.238\pm 0.019$ &$102.0\pm 1.1$&$-0.0026^{+0.0027}_{-0.0044}$  &-&-&$-369.588\pm0.310$ & $-2.610\pm0.330$&Moderate\\
$\Gamma_{b03}$ & $0.713^{+0.010}_{-0.013}$ & $0.239\pm 0.018$ &$101.9\pm 1.0$& $-0.00063^{+0.00055}_{-0.0063}$ &$-0.00063^{+0.00055}_{-0.0063}$ &-& $-370.008\pm0.054$&$-3.030\pm 0.124$&Moderate\\
$\Gamma_{b04}$ & $0.714^{+0.010}_{-0.013} $ & $0.238\pm 0.019$ &$102.0\pm 1.0$& $-0.0028^{+0.0029}_{-0.0040}$ & $-0.013\pm 0.098$&-& $-370.092\pm0.062$&$-3.114\pm0.128$&Moderate\\
$\Gamma_{c0}$ & $0.7100^{+0.0081}_{-0.011} $& $0.230\pm 0.019$&$102.32\pm 0.91$&$-0.021^{+0.069}_{-0.082}$&-&-& $-367.234\pm0.048$&$-0.256\pm0.122$&Inconclusive\\
$\Gamma_{d0}$ & $0.7139^{+0.0099}_{-0.013} $&$0.238\pm 0.019$ &$101.9\pm 1.0$& $0.0028^{+0.0040}_{-0.0028}$&-&-& $-369.822\pm0.221$&$-2.844\pm0.248$&Moderate\\
$\Gamma_{e0}$ &$0.7095^{+0.0078}_{-0.011} $& $0.230^{+0.029}_{-0.025} $&$102.33\pm 0.92$ &$-0.014^{+0.067}_{-0.076}$ &-&-&$-367.295\pm0.067$ &$-0.317\pm0.130$&Inconclusive\\
$\Gamma_{a1}$ &$0.713^{+0.010}_{-0.013}$ &$0.240\pm 0.023$ &$101.9\pm 1.1$ &$0.0027^{+0.0043}_{-0.0027}$ &-& $-0.006\pm 0.041$& $-368.923\pm0.070$ & $-1.945\pm0.132$&Weak\\
$\Gamma_{a2}$ & $0.7096^{+0.0083}_{-0.011}$& $0.236^{+0.039}_{-0.032}$& $102.24\pm 0.98 $&-&$-0.018^{+0.053}_{-0.062}  $ &$-0.003\pm 0.039$ &$-367.199\pm0.126$ & $-0.221\pm0.168$&Inconclusive\\
$\Gamma_{a3}$ & $0.713^{+0.010}_{-0.013}$&$0.243^{+0.020}_{-0.028}$ & $101.9\pm 1.0$ & $-0.0045^{+0.012}_{-0.0042}$&$-0.0045^{+0.012}_{-0.0042}$&$-0.006\pm 0.041 $ &$-368.145\pm0.776$ & $-1.167\pm0.784$&Weak\\
$\Gamma_{a4}$ & $0.713^{+0.011}_{-0.013}$ &$0.235\pm 0.037$ &$101.9\pm 1.1$ & $0.0026^{+0.0054}_{-0.0025}$&$0.017\pm 0.072$ &$-0.008\pm 0.041$ &$-369.101\pm0.181$ & $-2.123\pm0.213$&Weak\\
$\Gamma_{b1}$ & $0.713^{+0.010}_{-0.013}$ &$0.240\pm 0.023$ & $101.9\pm 1.0$&$-0.0027^{+0.0028}_{-0.0042}$ &-&$-0.006\pm 0.041$ & $-369.987\pm0.167$&$-3.009\pm 0.201$ &Moderate\\
$\Gamma_{b2}$ &$0.7095^{+0.0079}_{-0.011} $ &$0.226^{+0.015}_{-0.013}$ &$102.40\pm 0.90$ &-& $-0.005\pm 0.093$&$0.000\pm 0.035$ & $-367.399\pm0.030$&$-0.421\pm0.116$ &Inconclusive\\
$\Gamma_{b3}$ & $0.7134^{+0.0098}_{-0.013}$ &$0.240\pm 0.023$ & $101.9\pm 1.1$ &$-0.0027^{+0.0027}_{-0.0042}$ &$-0.0027^{+0.0027}_{-0.0042}$& $-0.006\pm 0.040$& $-370.194\pm0.036$&$-3.216\pm 0.118$ &Moderate\\
$\Gamma_{b4}$ & $0.7136^{+0.0099}_{-0.013} $&$0.240\pm 0.023$ &$101.9\pm 1.0$ & $-0.0029^{+0.0027}_{-0.0040}$& $-0.010\pm 0.096$&$-0.007\pm 0.040$ &$-370.220\pm0.044$ &$-3.242\pm 0.120$ &Moderate\\
$\Gamma_{c}$ &$0.7097^{+0.0081}_{-0.011} $ &$0.231\pm 0.021$ &$102.33\pm 0.90$ &$-0.021^{+0.070}_{-0.086}$ &-&	$-0.002\pm 0.037$&$-367.397\pm0.041$ & $-0.419\pm 0.119$&Inconclusive\\
$\Gamma_{d}$ &$0.7141^{+0.0098}_{-0.013} $ &$0.241\pm 0.023$ &$101.9\pm 1.0$ & $0.0029^{+0.0040}_{-0.0027}$&-&$-0.007\pm 0.041$ & $-369.436\pm0.528$&$-2.458\pm0.540$ &Weak\\
$\Gamma_{e}$ & $0.7089^{+0.0087}_{-0.011} $&$0.230^{+0.032}_{-0.026}$ &$102.30\pm 0.92$ &$-0.011^{+0.066}_{-0.076}  $ &-&$-0.001\pm 0.038$ & $-367.298\pm0.147$&$-0.32\pm0.185$ &Inconclusive\\
\hline
\end{tabular}}}
\end{table}
\end{landscape}

\section{\label{sec:finalremarks}Final Remarks}

Relaxing the conventional assumption of a purely gravitational interaction between the dark energy and dark matter components introduces substantial alterations in the predicted evolution of the universe. 
In this paper, we have performed a Bayesian model selection analysis to compare the observational viability of different classes of interacting models with the standard $\Lambda$CDM cosmology. We have found evidence disfavoring (weakly and moderately) about two-thirds  of the studied interacting models with respect to $\Lambda$CDM when a full joint analysis using JLA + BAO + CC + CMB data is considered. The remaining one-third, however, is able to provide a similar description to the data as the standard cosmology (see figure \ref{comparison}). Some interesting features have emerged in considering independently the full analysis using the standard BAO3 measurements and the full analysis using BAO2 measurements, which are more model independent than BAO3~\cite{Carvalho:2015ica}. In figure \ref{comparison} we see that in most of the cases the evidence disfavoring interacting models are weaker when we use BAO2 instead of BAO3 in the full joint analysis. As mentioned earlier, this result is expected since BAO3 data use $\Lambda$CDM as fiducial model, which may introduce a model-dependency in these data. Nevertheless, in considering separately scenarios \eqref{sc1} and \eqref{sc2} as priors we have noticed that the model-dependecy of BAO3 seems to be  partially contained in the estimation of $r_d$ as the sound horizon at the drag epoch. Finally, we emphasize that comparative analysis as the one performed in this paper is important to observationally select viable and non-viable classes of interacting scenarios. As summarized in figure \ref{comparison}, the current observational data are able to select between linear and non-linear interacting terms, showing a slightly preference for the latter type, i.e., among the twenty-one interacting scenarios shown in figure \ref{comparison} fifteen corresponds to a linear interaction and six to a non-linear interaction. From these, twelve linear scenarios are ruled out with weak or moderate evidence, meanwhile, only two of the non-linear scenarios are ruled out with weak or moderate evidence.\\

{\it Acknowledgements.} AC is partially supported by CONICYT through program Becas Chile de Postdoctorado en el Extranjero grant no. 74170121 and  Direcci\'on de Investigaci\'on Universidad del B\'io-B\'io through grant no. GI-172309/C. BS is supported by the DTI-PCI program of the Brazilian Ministry of Science, Technology, and Innovation (MCTI). TF thanks the financial support from CAPES/Brasil. JSA acknowledges support from CNPq (grants  no. 310790/2014-0  and  400471/2014-0)  and FAPERJ (grant no.  204282). AC would like to thank Javier Gonzalez for his assistance with JLA codes.

\bibliographystyle{apsrev4-1}
\bibliography{refs}

\appendix
\section{\label{sec:3BAO}BAO data}

In using BAO data shown in table \ref{TableBAOb} we have to consider some issues pointed out in the literature. For instance, in order to obtain the BAO signal from galaxy catalogues through the 2-point correlation function, it is necessary to assume a fiducial model in order to transform redshift into comoving distances~\cite{Percival:2009xn}. On the other hand, the fiducial model considered in the literature (usually the $\Lambda $CDM model) assumes different values for the cosmological parameters and consequently the derived $r_d^{\rm{fid}}$ value results to be different in each case. Besides, to compute $r_d^{\rm{fid}}$ authors use the Eisenstein's or CAMB formulas which could bring discrepancies in using these data combined to constrain cosmological models \cite{Mehta:2012hh}.

In table \ref{TableBAOb} we show the current available BAO measurements, here we have not included the data reported in references \cite{bao_anderson} or \cite{bao_blake} given that these measurements have been re-analyzed in references \cite{Cuesta:2015mqa} and \cite{Kazin:2014qga}, respectively. In these last references the authors apply the reconstruction of the baryonic acoustic feature technique \cite{Eisenstein2007, Padmanabhan:2012hf} to the former measurements. This reconstruction method leads to a more precise determination of the BAO signal, but the cost is to introduce model dependencies into the BAO data (see the discussion in reference~\cite{Anselmi:2018hdn}). In obtaining the results in section \ref{sec:results} we have used BOSS-LOWZ and BOSS-CMASS from reference \cite{Cuesta:2015mqa}, as well as eBOSS from reference \cite{Ata:2017dya}, nevertheless  these data were updated in references \cite{Alam:2016hwk} and \cite{Hou:2018yny}, respectively. For completeness we perform the analysis with an updated set of BAO data including the results in references \cite{Beutler:2011hx}, \cite{Ross:2014qpa}, \cite{Alam:2016hwk}, \cite{Hou:2018yny}  and \cite{Bourboux:2017cbm},  we show these results in table \ref{full_newBAO} of Appendix \ref{pantheon}.

Finally, it is worth to point it out that references \cite{Bourboux:2017cbm,Ata:2017dya,Hou:2018yny} are using quasars samples to detect the BAO signal at high redshifts whereas all the other measurements are performed from galaxy samples. As pointed out in reference \cite{Aghanim:2018eyx}, quasar Lyman $\alpha$ measurements \cite{Bourboux:2017cbm} require a number of additional assumptions, including universality of quasar continuum spectra, modelling of metal-line and high-column-density neutral hydrogen absorbers and spatial fluctuations in the UV ionizing flux, consequently we include this data only in the results shown in table \ref{full_newBAO} in Appendix \ref{pantheon}.

\begin{table}[h!]
\centering
\caption{\label{TableBAOb}The current BAO measurements available in the literature.}
\begin{tabular}{c|lllcc}
\hline\hline
Survey&$z_{\rm{eff}}$&Reported Parameter&Redshift Sample&$r_d^{\rm{fid}}$ [Mpc]& Ref. / Year\\\hline
6dFGS&0.106&$d_z^{-1}=0.3360\pm0.0150$& $0.00<z<0.24$&$-$&\cite{Beutler:2011hx} 2011 \\
WiggleZ&0.440&$d_z=(1716\pm83)/r_d^{\rm{fid}}$&$0.20<z<1.00$&148.60&\cite{Kazin:2014qga} 2014\\
WiggleZ&0.600&$d_z=(2221\pm101)/r_d^{\rm{fid}}$&$0.20<z<1.00$&148.60&\cite{Kazin:2014qga} 2014\\
WiggleZ&0.730&$d_z=(2516\pm86)/r_d^{\rm{fid}}$&$0.20<z<1.00$&148.60&\cite{Kazin:2014qga} 2014\\
SDSS-MGS&0.150&$d_z=(664\pm25   )/r_d^{\rm{fid}}$&$0.00<z<0.20$&148.69&\cite{Ross:2014qpa} 2015\\
BOSS-DR12&{0.380}&{$D_M\frac{r_d^{\rm{fid}}}{r_d}=(1512.39\pm24.99)$ } &{$0.2<z<0.75$}&{147.78}&{\cite{Alam:2016hwk} 2017}\\
BOSS-DR12&{0.380}&{$D_H/r_d=(81.21\pm2.37)/r_d^{\rm{fid}}$ } &{$0.2<z<0.75$}&{147.78}&{\cite{Alam:2016hwk} 2017}\\
BOSS-DR12&{0.510}&{$D_M\frac{r_d^{\rm{fid}}}{r_d}=(1975.22\pm30.10)$}&{$0.2<z<0.75$}&{147.78}&{\cite{Alam:2016hwk} 2017}\\
BOSS-DR12&{0.510}&{$D_H/r_d=(90.90\pm2.33)/r_d^{\rm{fid}}$ }&{$0.2<z<0.75$}&{147.78}&{\cite{Alam:2016hwk} 2017}\\
BOSS-DR12&{0.610}&{$D_M\frac{r_d^{\rm{fid}}}{r_d}=(2306.68\pm37.08)$} &{$0.2<z<0.75$}&{147.78}&{\cite{Alam:2016hwk} 2017}\\
BOSS-DR12&{0.610}&{$D_H/r_d=(98.96\pm2.50)/r_d^{\rm{fid}}$} &{$0.2<z<0.75$}&{147.78}&{\cite{Alam:2016hwk} 2017}\\
BOSS-Ly$\alpha$&2.400&${D_M/r_d=36.6\pm1.2}$&$2.00\le z\le 3.50$&147.33&\cite{Bourboux:2017cbm} 2017\\
BOSS-Ly$\alpha$&2.400&{$D_H/r_d=8.94\pm0.22$}&$2.00\le z\le 3.50$&147.33&\cite{Bourboux:2017cbm} 2017\\
eBOSS&1.520&$d_z=26.47\pm1.23$&$0.80<z<2.20$&147.78&\cite{Hou:2018yny} 2018\\
\hline\hline
\end{tabular}
\end{table}

\section{\label{pantheon}Pantheon compilation}
Recently the Pantheon compilation of SNe Ia was released \cite{Scolnic:2017caz}, the sample consists of 1048 SNe Ia spectroscopically confirmed in the redshift range $0.01<z<2.3$. The distance modulus of SNe Ia is calculated from the light curves using the empirical relation
\begin{equation}
\mu_B = m_B^* + M_B - \alpha \times x_1 + \beta \times c \,,
\end{equation}
where $m_B^*$ is the B-band apparent magnitude, $M_B$ is the absolute magnitude, $x_1$ and $c$ are the time stretching of the light curve and the supernovae color at its maximum brightness, respectively. Usually, the nuisance parameters $\alpha$ and $\beta$ are regarded as free parameters and they are constrained together with cosmological parameters (this was done with the JLA compilation). The Pantheon sample is calibrated using the BEAMS with Bias Corrections method \cite{Kessler:2016uwi}, in which the nuisance parameters $\alpha$ and $\beta$ are determined by fitting to a randomly chosen reference cosmology with the cosmological parameters fixed. In this sense, the reference \cite{Scolnic:2017caz} reported a corrected apparent magnitude for all the SNe and therefore, to calculate the distance modulus we do not need to consider the color and stretch corrections.

In tables \ref{full_PB2} and \ref{full_PB3} we show the results of considering the full joint analysis including the Pantheon sample (instead of JLA) with scenario \eqref{sc2} as priors. These results are analogous to those shown in tables \ref{full_nJB2} and \ref{full_nJB3}.
In comparing tables \ref{full_PB2} to \ref{full_nJB2} and \ref{full_PB3} to \ref{full_nJB3} we notice that while the inconclusive evidences remain all the same, the weak and moderate evidences fluctuates. Besides, the variation of the best fit for the better constrained parameters, $h$ and $r_d$, is always inside the 1$\sigma$ error estimation. 

The results in tables \ref{full_PB2} and \ref{full_PB3} have negligible changes in the best fit parameter estimation and the evidence when we consider $z_*$ fixed instead of deriving it from \eqref{zdec}, we have omitted showing these results for brevity.

In table \ref{full_newBAO} we show the full joint analysis (Pantheon+BAO+CC+CMB) with
the set of priors \eqref{sc2} and an updated set of BAO data, including data from:  6dFGS \cite{Beutler:2011hx}, SDSS-MGS \cite{Ross:2014qpa}, BOSS-DR12 \cite{Alam:2016hwk},  eBOSS \cite{Hou:2018yny} and BOSS-Ly$\alpha$ \cite{Bourboux:2017cbm}. In obtaining these results we have considered $z_*$ as derived from \eqref{zdec}. We notice that the differences in parameter estimation between tables \ref{full_PB3} and \ref{full_newBAO} are negligible in most of the cases and the major discrepancy is presented in the interpretation of the strength of the evidence, in those cases where the evidence is weak in table \ref{full_PB3} it becomes moderate in table \ref{full_newBAO}, in both cases the evidence is supporting $\Lambda$CDM.

\begin{landscape}
\begin{table}[h!]
\centering
\caption{\label{full_PB2}Best fit parameters for the joint analysis Pantheon + BAO2 + CC + CMB. The last three columns show the Bayesian evidence ($\ln \mathcal{E}$), the Bayes factor ($\ln B$) and the interpretation of the strength of the evidence. Note that $\ln B<-1$ favors the $\Lambda$CDM model. These results consider as priors scenario \eqref{sc1}.}
\resizebox{1.5\textwidth}{!}{%
{\renewcommand{\arraystretch}{1.5}%
\begin{tabular}{c|ccccccccc}\hline
Model & $h$ & $\Omega_{dm0}$  &$r_dh$ [Mpc]& $\alpha$ & $\beta$&$\gamma_x$&$\ln \mathcal{E}$&$\ln B$&Interpretation\\\hline
$\Lambda$CDM & $0.6991^{+0.0091}_{-0.011} $ & $0.238\pm 0.012$ &$105.6\pm 1.1$ &-&-&-&$-540.626\pm0.224$& -&\\
$\omega$CDM & $0.7000^{+0.0079}_{-0.011}$ & $0.234^{+0.016}_{-0.014}$ &$105.6\pm 1.0$ &-&-& $0.014^{+0.041}_{-0.034}$ & $-540.808\pm0.052$ & $-0.182\pm0.230$&Inconclusive\\
$\Gamma_{a01}$ & $0.710^{+0.010}_{-0.013}$& $0.257\pm 0.018$&$105.3\pm 1.0$& $0.0046^{+0.0037}_{-0.0020}$ &-&-&$-543.095\pm0.026$&  $-2.469\pm0.226$&Weak\\
$\Gamma_{a02}$ & $0.7055^{+0.0077}_{-0.011}$ & $0.264^{+0.031}_{-0.026}$ &$105.4\pm 1.0$&-& $-0.057^{+0.050}_{-0.057}$ &-& $-540.910\pm0.036$&$-0.284\pm 0.227$ &Inconclusive\\
$\Gamma_{a03}$ & $0.710\pm 0.012$ &$0.261^{+0.013}_{-0.026}$ &$105.2\pm 1.1$& $-0.0069^{+0.016}_{-0.0089}$ & $-0.0079^{+0.0064}_{-0.0047}$ &-&$-542.767\pm0.044$ &$-2.141\pm 0.228$ &Weak\\
$\Gamma_{a04}$ &$0.7105^{+0.0091}_{-0.012}$ & $0.256\pm 0.030$ &$105.3\pm 1.0$&$0.00297^{+0.0060}_{-0.00054}$ & $0.006^{+0.069}_{-0.078}$ &-& $-543.251\pm0.047$& $-2.625\pm0.229$&Moderate\\
$\Gamma_{b01}$ & $0.7116^{+0.0092}_{-0.012}$  & $0.257\pm 0.018 $ &$105.3\pm 1.1$&$-0.0052^{+0.0024}_{-0.0031}$  &-&-&$-543.005\pm0.088$ & $-2.379\pm0.241$&Weak\\
$\Gamma_{b03}$ & $0.711^{+0.010}_{-0.012}$ & $0.258\pm 0.018$ &$105.3\pm 1.1$& $-0.0001^{+0.0026}_{-0.0083}$ &$-0.0001^{+0.0026}_{-0.0083}$ &-& $-542.991\pm0.065$&$-2.365\pm0.233$&Weak\\
$\Gamma_{b04}$ & $0.7110^{+0.0094}_{-0.012}$ &$0.256^{+0.016}_{-0.018}$ &$105.3\pm 1.1$& $-0.0052\pm 0.0029$ & $-0.012\pm 0.092$&-& $-542.097\pm0.611$&$-1.471\pm0.651$&Weak\\
$\Gamma_{c0}$ & $0.7032^{+0.0066}_{-0.010}$& $0.246\pm 0.017$&$105.5\pm 1.0$&$-0.055^{+0.061}_{-0.081}$&-&-& $-540.744\pm0.175$&$-0.118\pm0.284$&Inconclusive\\
$\Gamma_{d0}$ & $0.7112^{+0.0094}_{-0.012}$&$0.257\pm 0.018$ &$105.3\pm 1.0$& $0.0051^{+0.0031}_{-0.0024}$&-&-& $-543.044\pm0.051$&$-2.418\pm0.230$&Weak\\
$\Gamma_{e0}$ &$0.7024^{+0.0074}_{-0.011}$ & $0.251^{+0.026}_{-0.022}$&$105.5\pm 1.0$ &$-0.049^{+0.070}_{-0.079}$ &-&-&$-540.898\pm0.044$ &$-0.272\pm0.228$&Inconclusive\\
$\Gamma_{a1}$ &$0.7116^{+0.0093}_{-0.012}$ & $0.257\pm 0.022$	&$105.2\pm 1.0$&$0.00320^{+0.0053}_{-0.00040}$ &-& $-0.003\pm 0.039$& $-543.144\pm0.057$&$-2.518\pm0.231$&Moderate\\
$\Gamma_{a2}$ & $0.7044^{+0.0085}_{-0.011} $& $0.262^{+0.035}_{-0.029}$&$105.4\pm 1.0$&-&$-0.055^{+0.049}_{-0.059}$ &$0.005^{+0.040}_{-0.036}$ &$-541.025\pm0.041$ &$-0.399\pm0.228$&Inconclusive\\
$\Gamma_{a3}$ & $0.710^{+0.010}_{-0.012}$&$0.258^{+0.021}_{-0.025} $ &$105.3\pm 1.0$&$0.0001^{+0.0084}_{-0.0024}$ & $0.0001^{+0.0084}_{-0.0024}$ &$-0.003\pm 0.041$ & $-542.909\pm0.121	$&$-2.283\pm0.254$&Weak\\
$\Gamma_{a4}$ &$0.7113^{+0.0095}_{-0.012}$ & $0.256\pm 0.033$ &$105.3\pm 1.0$& $0.0055^{+0.0037}_{-0.0029}$ & $0.008\pm 0.070$  &$-0.004\pm 0.040$ & $-543.137\pm0.145$ &$-2.511\pm0.267$&Moderate\\
$\Gamma_{b1}$ &$0.7116^{+0.0090}_{-0.012}$ & $0.257\pm 0.022$&$105.3\pm 1.0$&$-0.0053^{+0.0024}_{-0.0032}$ &-& $-0.002\pm 0.039$& $-543.205\pm0.031$&$-2.579\pm0.226$&Moderate\\
$\Gamma_{b2}$ & $0.7017^{+0.0074}_{-0.010}$ & $0.232^{+0.015}_{-0.013}$&$105.57\pm 0.99$&-& $0.000\pm 0.092$ & $0.016\pm 0.034$ &$-541.046\pm0.041$ &$-0.420\pm0.228$&Inconclusive\\
$\Gamma_{b3}$ &$0.7103^{+0.0090}_{-0.012}$ &$0.257\pm 0.022$ &$105.3\pm 1.0$& $-0.0050^{+0.0025}_{-0.0033}$&$-0.0050^{+0.0025}_{-0.0033}$& $-0.004\pm 0.040$& $-543.054\pm0.135$&$-2.428\pm0.262$&Weak\\
$\Gamma_{b4}$ & $0.7124^{+0.0075}_{-0.012}$ &$0.257\pm 0.022$ &$105.3\pm 1.0$&  $-0.0054^{+0.0026}_{-0.0030}$ & $-0.016\pm 0.093$ &$-0.003\pm 0.038$ & $-543.349\pm0.028$&$-2.723\pm0.226$&Moderate\\
$\Gamma_{c}$ &$0.7038^{+0.0071}_{-0.0098}$ &$0.241\pm 0.020$ &$105.5\pm 1.0$& $-0.049^{+0.066}_{-0.079}$ &-& $0.015^{+0.040}_{-0.035}$ & $-540.885\pm0.124$&$-0.259\pm0.256$&Inconclusive\\
$\Gamma_{d}$ & $0.7112^{+0.0088}_{-0.012}$ & $0.257\pm 0.023$ &$105.3\pm 1.1$& $0.0050^{+0.0032}_{-0.0023}$&-& $-0.003\pm 0.042$ & $-543.136\pm0.070$ &$-2.510 \pm 0.235$&Moderate\\
$\Gamma_{e}$ &$0.7034^{+0.0072}_{-0.010}$ & $0.243^{+0.030}_{-0.021}$ &$105.6\pm 1.0$& $-0.038^{+0.058}_{-0.077}$ 	&-& $0.013^{+0.041}_{-0.036}$ & $-541.109\pm0.043$ &$-0.483\pm0.228$&Inconclusive\\
\hline
\end{tabular}}}
\end{table}
\end{landscape}

\begin{landscape}
\begin{table}[ht!]
\centering
\caption{\label{full_PB3}Best fit parameters for the joint analysis Pantheon + BAO3 + CC + CMB. The last three columns show the Bayesian evidence ($\ln \mathcal{E}$), the Bayes factor ($\ln B$) and the interpretation of the strength of the evidence. Note that $\ln B<-1$ favors the $\Lambda$CDM model. These results consider as priors scenario \eqref{sc2}.}
\resizebox{1.5\textwidth}{!}{%
{\renewcommand{\arraystretch}{1.5}%
\begin{tabular}{c|ccccccccc}\hline
Model & $h$ & $\Omega_{dm0}$  &$r_dh$ [Mpc]& $\alpha$ & $\beta$&$\gamma_x$&$\ln \mathcal{E}$&$\ln B$&Interpretation\\\hline
$\Lambda$CDM & $0.7038^{+0.0081}_{-0.010} $ & $0.232^{+0.012}_{-0.010}$ &$102.13\pm 0.89$ &-&-&-&$-532.642\pm0.022$& -&\\
$\omega$CDM & $0.7030^{+0.0087}_{-0.010} $ & $0.233\pm 0.015$ &$102.07\pm 0.90$ &-&-& $0.000\pm 0.038$ & $-532.511\pm0.215$ & $0.131\pm0.216$&Inconclusive\\
$\Gamma_{a01}$ & $0.713^{+0.011}_{-0.013}$& $0.244\pm 0.016$ & $101.70\pm 0.97$& $0.0033^{+0.0038}_{-0.0023}$&-&-&$-535.032\pm0.315$& $-2.390\pm 0.316$ &Weak\\
$\Gamma_{a02}$ & $0.7051^{+0.0085}_{-0.011} $ & $0.240^{+0.032}_{-0.029}$ &$102.01\pm 0.92$&-& $-0.016^{+0.054}_{-0.061}$ &-& $-532.780\pm0.150$&$-0.138\pm0.152$&Inconclusive\\
$\Gamma_{a03}$ & $0.712^{+0.010}_{-0.013}$ &$0.244\pm 0.017$ &$101.67\pm 0.95$& $0.0019^{+0.0054}_{-0.0010}$& $0.0019^{+0.0054}_{-0.0010}$ &-&$-535.264\pm0.075$ &$-2.622\pm0.078$ &Moderate\\
$\Gamma_{a04}$ &$0.712^{+0.010}_{-0.013}$ & $0.231\pm 0.030$ &$101.70\pm 0.94$ &$0.0047^{+0.0039}_{-0.0029}$ & $0.040\pm 0.070$ &-& $-535.535\pm0.026$& $-2.893\pm0.034$&Moderate\\
$\Gamma_{b01}$ & $0.713^{+0.011}_{-0.013}$  & $0.244\pm 0.016$ &$101.70\pm 0.96 $&$-0.0037^{+0.0026}_{-0.0035}$&-&-&$-535.283\pm0.044$ & $-2.641\pm0.049$&Moderate\\
$\Gamma_{b03}$ & $0.713^{+0.010}_{-0.013}$ & $0.245\pm 0.016$ &$101.69\pm 0.96$& $-0.0033^{+0.0021}_{-0.0039}$ &$-0.0033^{+0.0021}_{-0.0039}$ &-& $-534.834\pm0.329$&$-2.192\pm0.330$&Weak\\
$\Gamma_{b04}$ & $0.7137^{+0.0093}_{-0.012} $ &$0.244\pm 0.016$ & $101.74\pm 0.96$& $-0.0038^{+0.0027}_{-0.0032}$ & $-0.012\pm 0.096$&-& $-535.456\pm0.040$&$-2.814\pm0.046$&Moderate\\
$\Gamma_{c0}$ & $0.7043^{+0.0083}_{-0.011} $& $0.237\pm 0.017$&$102.06\pm 0.88$&$-0.019\pm 0.075$&-&-& $-532.699\pm0.081$&$-0.057\pm0.084$&Inconclusive\\
$\Gamma_{d0}$ & $0.7133^{+0.0098}_{-0.013} $&$0.248\pm 0.020$ &$101.69\pm 0.98$& $0.0038^{+0.0033}_{-0.0026}$&-&-& $-535.411\pm0.027$&$-2.769\pm0.035$&Moderate\\
$\Gamma_{e0}$ &$0.7041^{+0.0082}_{-0.010}$ & $0.234^{+0.026}_{-0.023}$&$102.07\pm 0.88 $ &$-0.007\pm 0.071$ &-&-&$-532.835\pm0.042$ &$-0.193\pm0.047$&Inconclusive\\
$\Gamma_{a1}$ &$0.7133^{+0.0098}_{-0.012} $ & $0.248\pm 0.020$	&$101.70\pm 0.94$ &$0.0031^{+0.0043}_{-0.0017}$&-& $-0.012\pm 0.039$& $-535.455\pm0.024$&$-2.813\pm0.032$&Moderate\\
$\Gamma_{a2}$ & $0.7059^{+0.0079}_{-0.010} $& $0.242\pm 0.033$&$102.04\pm 0.91$ &-&$-0.021\pm 0.054$ &$-0.004\pm 0.038$ &$-533.200\pm0.060$&$-0.558\pm0.064$&Inconclusive\\
$\Gamma_{a3}$ & $0.713^{+0.010}_{-0.012}$&$0.248\pm 0.020$ &$101.70\pm 0.94$&$0.0033^{+0.0043}_{-0.0018}$ & $0.0033^{+0.0043}_{-0.0018}$ &$-0.014\pm 0.040$ & $-535.054\pm0.255	$&$-2.412\pm0.256$&Weak\\
$\Gamma_{a4}$ &$0.714^{+0.010}_{-0.012}$ & $0.235\pm 0.034$ &$101.74\pm 0.91$& $0.0048^{+0.0039}_{-0.0026}$ & $0.035\pm 0.068$  &$-0.013\pm 0.039$ & $-535.340\pm0.139$ &$-2.698\pm0.141$&Moderate\\
$\Gamma_{b1}$ &$0.7135^{+0.0094}_{-0.012}$ & $0.249\pm 0.020$&$101.70\pm 0.94$ &$-0.0041^{+0.0024}_{-0.0033}$ &-& $-0.015\pm 0.039$& $-535.164\pm0.238$&$-2.522\pm0.239$&Moderate\\
$\Gamma_{b2}$ & $0.7050^{+0.0078}_{-0.010} $ & $0.231^{+0.015}_{-0.012}$&$102.14\pm 0.87$&-& $-0.005\pm 0.094$ & $0.002\pm 0.035$ &$-532.765\pm0.049$ &$-0.123\pm0.054$&Inconclusive\\
$\Gamma_{b3}$ &$0.713^{+0.011}_{-0.012} $ &$0.249\pm 0.020$ &$101.73\pm 0.94$& $-0.0039^{+0.0024}_{-0.0034}$&$-0.0039^{+0.0024}_{-0.0034}$ & $-0.015\pm 0.040$& $-535.496\pm0.019$&$-2.854\pm0.029$&Moderate\\
$\Gamma_{b4}$ & $0.7145^{+0.0089}_{-0.012}$ &$0.248\pm 0.020$ &$101.75\pm 0.92$&  $-0.0043^{+0.0026}_{-0.0031}$ & $-0.014\pm 0.091$ &$-0.015\pm 0.038$ & $-535.518\pm0.064$&$-2.876\pm0.068$&Moderate\\
$\Gamma_{c}$ &$0.7049^{+0.0079}_{-0.010} $ &$0.236\pm 0.020$ &$102.06\pm 0.91$& $-0.019^{+0.070}_{-0.082}$ &-& $-0.002\pm 0.036$ & $-532.857\pm0.071$&$-0.215\pm0.074$&Inconclusive\\
$\Gamma_{d}$ & $0.713^{+0.010}_{-0.012}$ &$0.248\pm 0.020 $ &$101.70\pm 0.95$& $0.0038^{+0.0034}_{-0.0025}$&-& $-0.013\pm 0.040$ & $-535.327\pm0.074$ &$-2.685\pm0.077$&Moderate\\
$\Gamma_{e}$ &$0.7040^{+0.0083}_{-0.010} $ & $0.236^{+0.030}_{-0.024}$ &$102.08^{+0.84}_{-0.94}$& $-0.011\pm 0.069$ &-& $-0.002\pm 0.038$ & $-532.964\pm0.043$ &$-0.322\pm0.048$&Inconclusive\\
\hline
\end{tabular}}}
\end{table}
\end{landscape}

\begin{landscape}
\begin{table}[ht!]
\centering
\caption{\label{full_newBAO}Best fit parameters for the joint analysis Pantheon + BAO + CC + CMB. The last three columns show the Bayesian evidence ($\ln \mathcal{E}$), the Bayes factor ($\ln B$) and the interpretation of the strength of the evidence. Note that $\ln B<-1$ favors the $\Lambda$CDM model. These results consider as priors scenario \eqref{sc2} and an updated version of BAO data.}
\resizebox{1.5\textwidth}{!}{%
{\renewcommand{\arraystretch}{1.5}%
\begin{tabular}{c|ccccccccc}\hline
Model & $h$ & $\Omega_{dm0}$  &$r_dh$ [Mpc]& $\alpha$ & $\beta$&$\gamma_x$&$\ln \mathcal{E}$&$\ln B$&Interpretation\\\hline
$\Lambda$CDM & $0.7010\pm 0.0076$ & $0.2358\pm 0.0092$ &$102.14\pm 0.8$ &-&-&-&$-538.586\pm0.144$& -&\\
$\omega$CDM & $0.7014^{+0.0065}_{-0.0079}$ & $0.234\pm 0.010$ &$102.13^{+0.80}_{-0.91}$ &-&-& $0.006^{+0.036}_{-0.032}$ & $-539.018\pm0.018$ & $-0.432\pm 0.145$ &Inconclusive\\
$\Gamma_{a01}$ & $0.711^{+0.011}_{-0.013}$& $0.242\pm 0.011$ & $101.82\pm 0.88$& $0.0031^{+0.0030}_{-0.0026}$&-&-&$-541.591\pm0.061$& $-3.005\pm 0.156$ &Moderate\\
$\Gamma_{a02}$ & $0.7012^{+0.0064}_{-0.0079}$ & $0.248^{+0.030}_{-0.026}$ &$101.99\pm 0.88$&-& $-0.022\pm 0.047$ &-& $-539.436\pm0.018$&$-0.850\pm 0.145$&Inconclusive\\
$\Gamma_{a03}$ & $0.713^{+0.010}_{-0.013}$ &$0.240\pm 0.011$ &$101.87\pm 0.91$& $0.0036^{+0.0029}_{-0.0026}$& $0.0036^{+0.0029}_{-0.0026}$ &-&$-541.405\pm0.154$ &$-2.819\pm 0.211$ &Moderate\\
$\Gamma_{a04}$ &$0.712^{+0.010}_{-0.013}$ & $0.248\pm 0.028$ &$101.77\pm 0.93$ &$0.0032^{+0.0029}_{-0.0026}$ & $-0.011\pm 0.049$ &-& $-542.188\pm0.057$& $-3.602\pm0.155$&Moderate\\
$\Gamma_{b01}$ & $0.712^{+0.011}_{-0.013}$  & $0.242\pm 0.011$ &$101.85\pm 0.90$&$-0.0032^{+0.0026}_{-0.0030}$&-&-&$-541.632\pm0.068$ &$-3.046\pm 0.159$&Moderate\\
$\Gamma_{b03}$ & $0.711^{+0.011}_{-0.013}$ & $0.242\pm 0.011$ &$101.83\pm 0.89$& $-0.0031^{+0.0026}_{-0.0030}$ &$-0.0031^{+0.0026}_{-0.0030}$ &-& $-541.332\pm0.214$&$-2.746\pm 0.258$&Moderate\\
$\Gamma_{b04}$ & $0.7119^{+0.0096}_{-0.012} $ &$0.242\pm 0.011$ & $101.85\pm 0.89$& $-0.0033\pm 0.0027$ & $-0.013\pm 0.096$&-& $-541.647\pm0.062$&$-3.061\pm 0.157$&Moderate\\
$\Gamma_{c0}$ & $0.7013^{+0.0066}_{-0.0078}$& $0.244\pm 0.018$&$102.04\pm 0.85$&$-0.035\pm 0.067$&-&-& $-538.812\pm0.154$&$-0.226\pm 0.211$&Inconclusive\\
$\Gamma_{d0}$ & $0.7130^{+0.0097}_{-0.012} $&$0.242\pm 0.011$ &$101.83\pm 0.90$& $0.0035^{+0.0027}_{-0.0024}$&-&-& $-541.646\pm0.027$&$-3.060\pm 0.146$&Moderate\\
$\Gamma_{e0}$ &$0.7014^{+0.0063}_{-0.0078}$& $0.239^{+0.027}_{-0.022}$&$102.10\pm 0.85$ &$-0.014\pm 0.068 $ &-&-&$-538.970\pm0.062$ &$-0.384\pm 0.157$&Inconclusive\\
$\Gamma_{a1}$ &$0.712^{+0.010}_{-0.012}$ & $0.241\pm 0.013$	&$101.81\pm 0.92$ &$0.0034^{+0.0029}_{-0.0025}$&-& $0.005^{+0.038}_{-0.033}$& $-541.872\pm0.031$&$-3.286\pm 0.147$&Moderate\\
$\Gamma_{a2}$ & $0.7019^{+0.0060}_{-0.0074}$& $0.248^{+0.035}_{-0.031}$&$102.05\pm 0.86$ &-&$-0.024\pm 0.051$ &$-0.004\pm 0.036$ &$-539.407\pm0.085$&$-0.821\pm0.167$&Inconclusive\\
$\Gamma_{a3}$ & $0.712^{+0.010}_{-0.013}$&$0.239\pm 0.012$ &$101.78\pm 0.89$&$0.0033\pm 0.0027$ & $0.0033\pm 0.0027$ &$0.006\pm 0.035$ & $-541.674\pm0.139	$&$-3.088\pm 0.200$&Moderate\\
$\Gamma_{a4}$ &$0.7127^{+0.0095}_{-0.012} $ & $0.249^{+0.037}_{-0.032}$ &$101.75\pm 0.90$& $0.0035\pm 0.0026$ & $-0.013\pm 0.054$ &$-0.002\pm 0.039$ & $-542.408\pm0.025$ &$-3.822\pm 0.146$&Moderate\\
$\Gamma_{b1}$ &$0.7128^{+0.0095}_{-0.013} $ & $0.240\pm 0.012$&$101.82\pm 0.90$ &$-0.0034\pm 0.0026 $ &-& $0.004\pm 0.034$& $-541.900\pm0.035$&$-3.314\pm 0.148$&Moderate\\
$\Gamma_{b2}$ & $0.7016^{+0.0061}_{-0.0074}$ & $0.234\pm 0.010$&$102.16^{+0.78}_{-0.87}$&-& $-0.0095\pm 0.096$ & $0.004^{+0.036}_{-0.032}$ &$-538.976\pm0.070$ &$-0.390\pm 0.160$&Inconclusive\\
$\Gamma_{b3}$ &$0.7130^{+0.0099}_{-0.012} $ &$0.241^{+0.012}_{-0.013} $ &$101.83\pm 0.90 $& $-0.0035\pm 0.0027$&$-0.0035\pm 0.0027$ & $0.004\pm 0.035$& $-541.784\pm0.067$&$-3.198\pm 0.159$&Moderate\\
$\Gamma_{b4}$ & $0.7132^{+0.0093}_{-0.012} $ &$0.241\pm 0.012$ &$101.79\pm 0.88$&  $-0.0036\pm 0.0026$ & $-0.015\pm 0.092 $ &$0.006\pm 0.033$ & $-541.722\pm0.204$&$-3.136\pm 0.250$&Moderate\\
$\Gamma_{c}$ &$0.7014^{+0.0063}_{-0.0075}$ &$0.243\pm 0.020$ &$102.06\pm 0.85$& $-0.033^{+0.063}_{-0.075}$ &-& $-0.001^{+0.038}_{-0.034}$ & $-538.958\pm0.065$&$-0.372\pm0.158$&Inconclusive\\
$\Gamma_{d}$ & $0.7126^{+0.0095}_{-0.012} $ &$0.242\pm 0.013$ &$101.79\pm 0.89$&$0.0035\pm 0.0026$&-& $0.004\pm 0.034$ & $-541.914\pm0.025$ &$-3.328\pm 0.146$&Moderate\\
$\Gamma_{e}$ &$0.7012^{+0.0065}_{-0.0077}$ & $0.236^{+0.030}_{-0.023}$ &$102.12^{+0.80}_{-0.90}$& $-0.007\pm 0.065$ &-& $0.003\pm 0.035$ & $-539.191\pm0.080$ &$-0.605\pm 0.165$&Inconclusive\\
\hline
\end{tabular}}}
\end{table}
\end{landscape}
\section{\label{nuisance2}Nuisance parameters of the JLA compilation of SNe Ia}
\begin{figure}[ht!]
\centering
\includegraphics[scale=0.45]{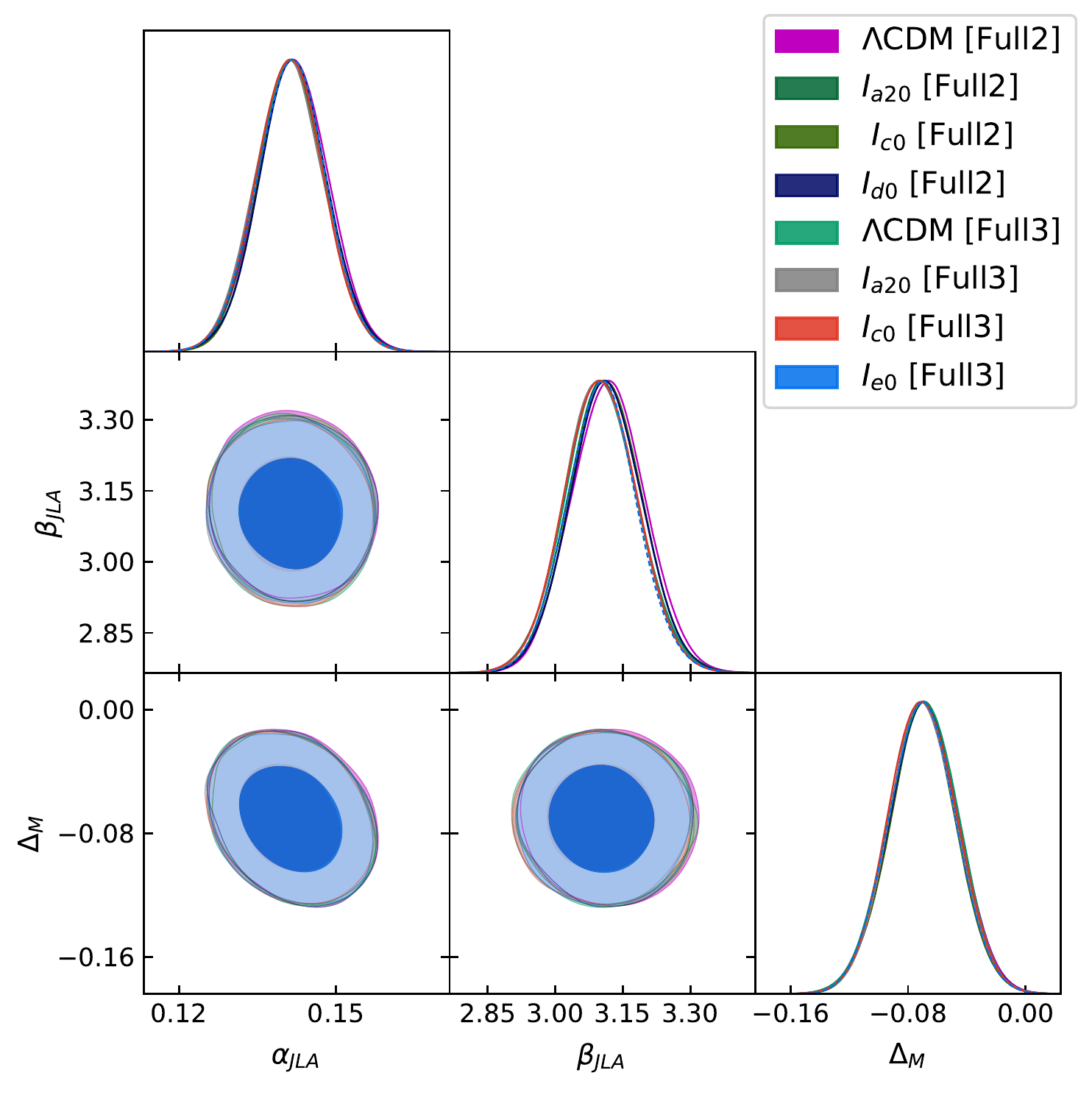}
\includegraphics[scale=0.45]{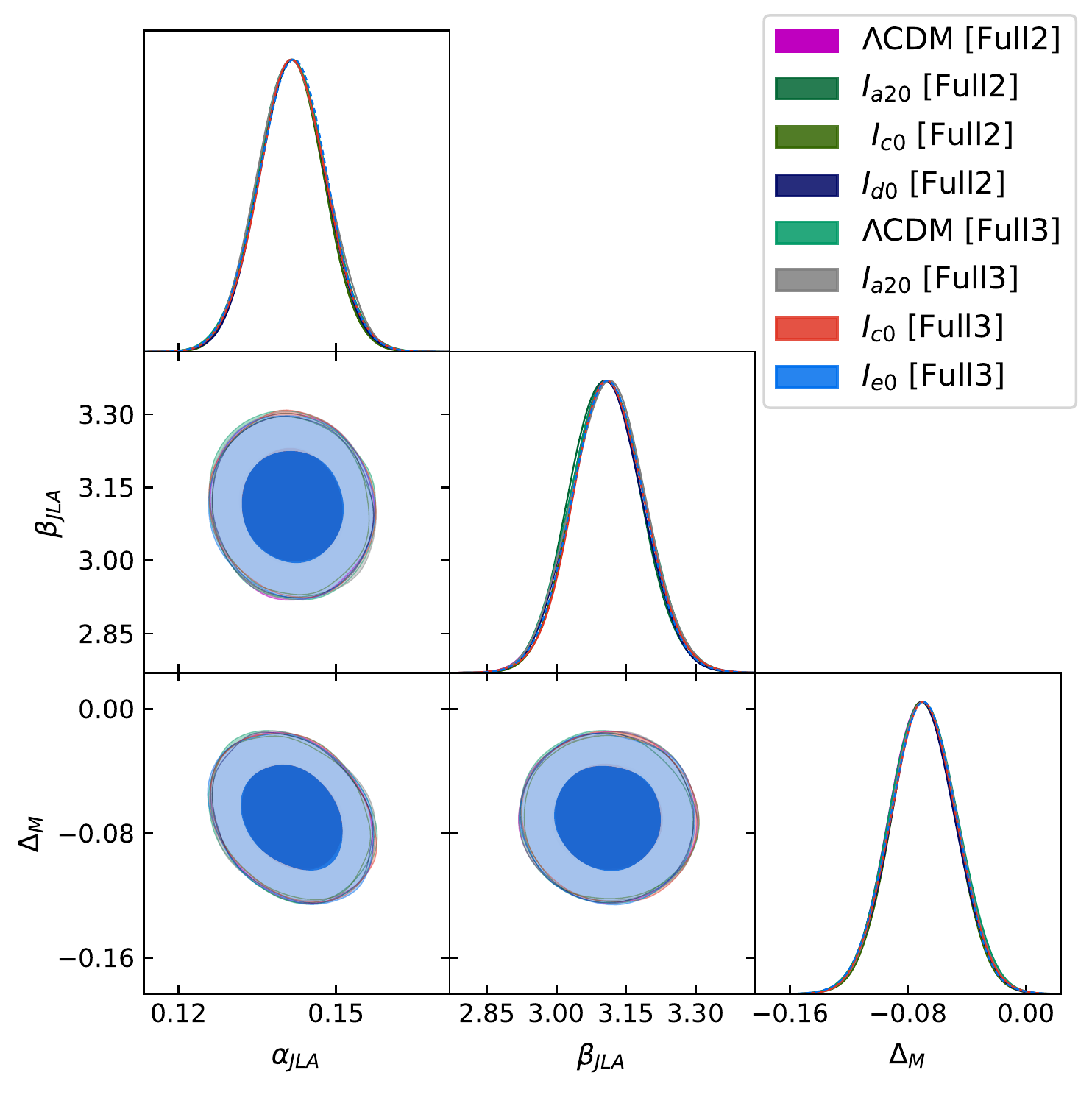}
\caption{\label{nuisance} Contour plots and PDFs for the nuisance parameters $\{\alpha_{JLA}\,,\beta_{JLA}\,,\Delta_M\}$ for some of the studied models, namely, $\Lambda$CDM, $I_{a20}$, $I_{c0}$ and $I_{e0}$. Full2 and Full3 represent respectively, the full joint analysis for JLA+BAO2+CC+CMB and JLA+BAO3+CC+CMB. The left and right panels show the results in using as priors scenario \eqref{sc1} and \eqref{sc2}, respectively.}
\end{figure}

In figure \ref{nuisance} we verify that there is no correlation among interacting parameters and the nuisance parameters of the JLA compilation, $\alpha_{JLA}$, $\beta_{JLA}$, $\Delta_M$. We show some of the studied models for brevity because there is no noticeable variation for other interacting scenarios. Our results indicate that the nuisance parameters are almost unaffected by the interacting scenarios.

\end{document}